\documentclass[acmsmall,screen]{acmart}

\AtBeginDocument{%
  }
\usepackage{makecell}
\usepackage{multirow}
\usepackage{tabularx}
\usepackage{soul}
\usepackage{graphicx}  
\usepackage{subcaption}
\usepackage[font=small,labelfont=bf]{caption}
\usepackage{booktabs}    
\usepackage{array}       

\begin{document}




\title{\textit{Eye2Recall}: Fusing Gaze and LLMs for Mixed-Initiative Reminiscence with Older Adults}


\author{Lei Han}
\orcid{0009-0001-7157-8702}
\affiliation{%
  \institution{Computational Media and Arts, The Hong Kong University of Science and Technology (Guangzhou)}
  \city{Guangzhou}
  \state{Guangdong}
  \country{China}}
  \email{lhan229@connect.hkust-gz.edu.cn}

\author{Mingnan Wei}
\orcid{0000-0003-1539-4420}
\affiliation{%
  \institution{Bioscience and Biomedical Engineering, The Hong Kong University of Science and Technology (Guangzhou)}
  \city{Guangzhou}
  \state{Guangdong}
  \country{China}}
  \email{mwei881@connect.hkust-gz.edu.cn}

\author{Qiongyan Chen}
\orcid{0009-0004-0893-7014}
\affiliation{%
  \institution{Computational Media and Arts, The Hong Kong University of Science and Technology (Guangzhou)}
  \city{Guangzhou}
  \state{Guangdong}
  \country{China}}
  \email{qchen580@connect.hkust-gz.edu.cn}

\author{Anqi Wang}
\orcid{0000-0003-4238-6581}
\affiliation{%
  \institution{Emerging Interdisciplinary Areas, Hong Kong University of Science and Technology}
 \city{Hong Kong}
  \country{China}}
\email{awangan@connect.ust.hk}

\author{Rong Pang}
\orcid{0009-0005-8127-9115}
\affiliation{%
  \institution{Edinburgh College of Art, University of Edinburgh}
 \city{Edinburgh}
  \country{United Kingdom}}
  \email{rong.pang2024@outlook.com}

\author{Kefei Liu}
\orcid{}
\affiliation{%
  \institution{Bioscience and Biomedical Engineering, The Hong Kong University of Science and Technology (Guangzhou)}
  \city{Guangzhou}
  \state{Guangdong}
  \country{China}}
  \email{kefeiliu@hkust-gz.edu.cn}

\author{Rongrong Chen}
\orcid{}
\affiliation{%
  \institution{Guangdong Provincial Key Laboratory IRADS, BNU-HKBU United International College}
 \city{Zhuhai}
  \country{China}}
  \email{ainerrchen@uic.edu.cn}

\author{David Yip}
\authornote{Corresponding Author}
\orcid{0000-0002-1745-4741}
\affiliation{%
  \institution{Computational Media and Arts, The Hong Kong University of Science and Technology (Guangzhou)}
  \city{Guangzhou}
  \state{Guangdong}
  \country{China}}
  \email{daveyip@hkust-gz.edu.cn}



\begin{abstract}

Photo-based reminiscence can support well-being in older adults, yet most systems remain text-driven and offer little real-time adaptivity. 
We first conduct expert interviews to derive design considerations for accessibility, cultural fit, and safe emotional engagement. We then implemented \textit{Eye2Recall}, an intelligent conversational interface that converts users’ gazes on old photos into mixed-initiative prompts for a large language model (LLM). We evaluated it in a pilot study with 12 older adults. Participants reported low-effort, smooth interactions, and perceived the agent’s questions as aligned with what they were looking at. Immediately after use, self-reported positive mood increased and negative mood decreased. Interviews further indicated that gaze-driven prompts helped retrieve concrete details and supported reflective storytelling. Our contribution is a concrete mechanism for gaze-to-prompt adaptivity that operationalizes mixed-initiative dialogue for older adults’ reminiscence experience.

\end{abstract}






\maketitle 

\section{Introduction}

\begin{figure*}[t]
  \centering
    \includegraphics[height=8cm]{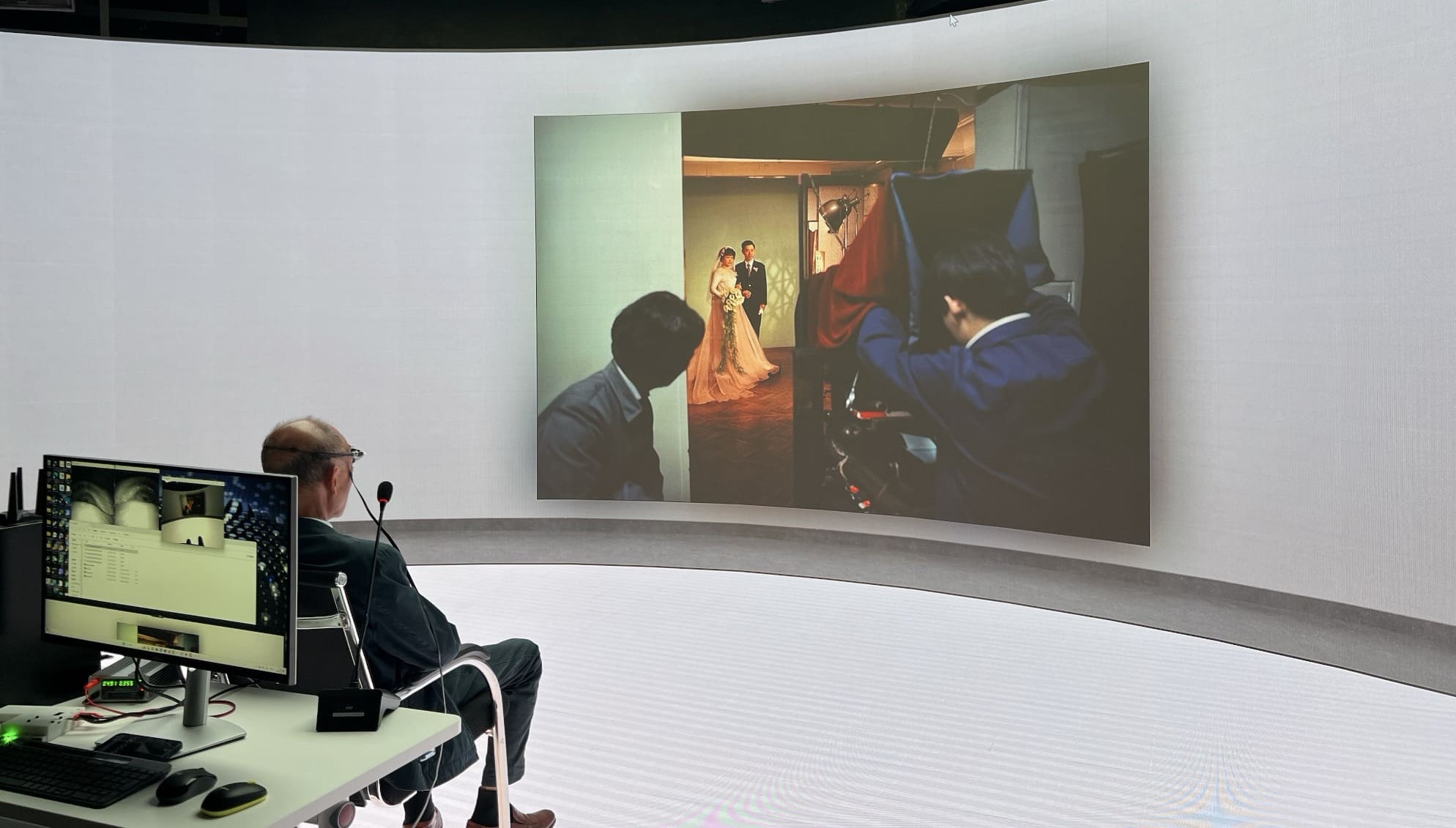}
  \caption{Participants use the \textit{Eye2Recall} prototype to interact with the LLM-powered agent.}
  \Description{Participants use the \textit{Eye2Recall} prototype to interact with the LLM-powered agent.}
  \label{fig:teaser}
\end{figure*}

Reminiscence is considered an effective approach for improving the well-being of older adults \cite{madoglou2017representations, bohlmeijer2007effects}, positively impacting emotional health for both healthy older adults and individuals with dementia \cite{fleury2022feeling, bryant2005using}. In addition, reminiscence can enhance communication skills and strengthen intergenerational relationships \cite{webster2007reminiscence, flottemesch2013learning, li2023exploring}. 

In recent decades, research has increasingly explored how technology can support reminiscence and enrich older adults’ daily lives. Various digital tools have been developed to evoke memories through interactive engagement with photos \cite{webster2007reminiscence, petrelli2008autotopography, kim2006cherish}, sounds \cite{jayaratne2016memory, dib2010sonic, jin2024exploring}, and other sensory stimuli. With the advancement of large language models (LLMs), studies have investigated LLM-powered solutions for digital reminiscence, including collecting personal memories \cite{peesapati2010pensieve}, supporting storytelling \cite{chu2016understanding}, and facilitating reminiscence conversations \cite{RW02}.

Despite these advances, existing technologies still face challenges in effectively supporting older adults’ reminiscence \cite{10.1145/3711094, yang2024talk2care}. Many LLM-powered conversational agents (CAs) rely primarily on text or speech inputs, which may create barriers due to complex user interfaces or literacy requirements \cite{stigall2019older, yang2024talk2care}. Additionally, these systems often lack personalization and struggle to adapt dynamically to users’ needs, while speech-based agents face difficulties understanding older adults’ accents and nuances, potentially reducing conversational coherence \cite{myers2018patterns, kim2024understanding}.

Previous studies suggest that eye-tracking technology can help older adults express their needs and preferences more effectively, particularly for those with communication difficulties \cite{pavic2021age, liu2021effect}. Recent HCI and AI research has employed gaze information to guide AI models by aligning attention with human fixations and areas of interest \cite{yan2024voila, salamatian2025chartgaze, lopez2024seeing, unger2025stare, trajkovska2024gaze2aoi}. However, little is known about using eye-tracking regions of interest (ROI) as cues for LLMs to support \textcolor{black}{gaze-driven} reminiscence dialogues with older adults. \textcolor{black}{Gaze information reflects pre-intentional attention dynamics and implicit cues~\mbox{\cite{10.5555/216164.216178,10.1145/3015783.3015794}}, which can support nuanced and unstructured interactive activities, such as reminiscence. Leveraging this capability, gaze information} allows conversational systems to adapt dynamically to users’ attention and interests, enhancing personalization and engagement in the such experiences. 
In this context, we propose two research questions:


\begin{itemize}
  \item \textbf{RQ1:} \textit{How can the integration of LLM with eye-tracking technology enhance the conversational experience for older adults during reminiscence activities?}
  
  \item \textbf{RQ2:} \textit{What effects does LLM-powered conversation with eye-tracking augmentation have on older adults’ workload, usability, and affect in photo-based reminiscence?}

\end{itemize}

To address these questions, we conducted a semi-structured interview study with experts to understand older adults’ needs and preferences during photo-based reminiscence. Based on the findings, we designed \textit{Eye2Recall}, an LLM-powered prototype that integrates eye-tracking with visual interest detection to adapt dialogue content in real time (\textcolor{black}{see Fig.~\ref{fig:teaser}}). To evaluate the prototype and derive design implications, we conducted a user study (N=12) aged 60 and above, including pre-evaluation, prototype testing, semi-structured interviews, and post-evaluation, to comprehensively investigate older adults’ reminiscence experiences with a gaze-augmented conversational system.
In summary, this work makes three key contributions:

\begin{itemize}
  \item Conducted expert interviews to identify design considerations, challenges, and opportunities for LLM tools that support photo-based reminiscence with older adults.
  \item Developed \textit{Eye2Recall}, a prototype that uses gaze-based cues to personalize LLM-mediated reminiscence conversations around old photos.
  \item Ran a user study with older adults to evaluate \textit{Eye2Recall}’s usability, workload, and user experience, yielding design insights and implications for future LLM-supported reminiscence systems.
\end{itemize}

\section{Related Work}
\subsection{Psychological Support with Reminiscence}
Reminiscence, which involves recalling and sharing past experiences, has long been recognized as a meaningful way to promote psychological well-being in older adults \cite{madoglou2017representations, bohlmeijer2007effects}. It can enhance mood, foster positive emotions, and build psychological resilience \cite{1.1, 2.2}. Such recollection is often triggered by sensory cues, such as photographs or music, which evoke autobiographical memories and shape corresponding emotional responses \cite{astell2010stimulating, tokunaga2021dialogue, LI2025100805, romaniuk1981looking}. However, access to human-facilitated reminiscence remains uneven: caregivers’ time is limited and professional services are costly, placing emotional and logistical burdens on families and care systems \cite{loureiro2013burden}. These constraints highlight the need for scalable, low-effort digital tools that can support memory elicitation in everyday contexts.

Beyond short-term emotional outcomes, reminiscence also strengthens self-identity and a sense of belonging by prompting reflection on life accomplishments and enabling the sharing of personal stories within social contexts \cite{2.4, lewis1971reminiscing}. As a therapeutic approach, reminiscence and narrative techniques have been used to engage people living with depression, dementia, or cognitive decline, often producing measurable improvements in emotional health \cite{fleury2022feeling, creative, r2.1}. Among different modalities, photo-elicited reminiscence has proven particularly effective: old images can cue autobiographical memory, help retrieve self-defining experiences, and reinforce a coherent sense of life continuity \cite{pinquart2012effects, tan2023effects, davison2016personalized, toledano2023effects}. This body of work motivates designing technologies that make photo-based reminiscence more accessible, adaptive, and emotionally supportive for older adults.

\subsection{Conversational Systems for Reminiscence}
Recent HCI research has increasingly explored how conversational agents (CAs) can facilitate reminiscence for older adults \cite{2.2.11, wu2019interactive, 10.1145/3708557.3716362, 10.1145/3180308.3180330}. Advances in natural language processing (NLP) and machine learning (ML) have enabled CAs capable of engaging users in fluid, human-like dialogue \cite{laranjo2018conversational}. For example, \textit{GoodTimes} is an interactive photo album application that uses AI to converse about personal pictures and life events, promoting engagement around family and social memories \cite{wangpromoting}. Such systems are commonly categorized as either task-oriented or open-domain dialogue systems \cite{10.1145/3383123, 10.1145/3678698.3687186}.

Task-oriented systems are often designed for therapeutic goals such as emotional regulation or cognitive intervention~\cite{10.1145/3172944.3173010, jin2024exploring}. \textit{Yeonheebot}, for instance, provides automated reminiscence therapy to mitigate depression and anxiety in older adults \cite{ryn2020yeonheebot}, demonstrating the potential of conversational technologies in dementia care and mental health contexts \cite{lazar2014systematic}. Other efforts have used NLP and ML to analyze older adults’ natural conversation patterns and promote social reminiscence \cite{yordanova2019automatic, ferrario2020social}. However, many of these systems remain rule-based or rely on predefined conversation flows, which limit their ability to handle unexpected topics or follow users’ emotional trajectories \cite{10.1145/3290605.3300932, 10.1145/3708359.3712091, lyu2025visual}.

Open-domain dialogue systems offer greater flexibility, supporting natural and wide-ranging conversation. Studies have shown that open-ended questions can improve autobiographical recall \cite{brainsci12030374}, but generating such prompts in a way that feels empathetic and contextually appropriate remains a key challenge. The rapid advancement of large language models (LLMs) has transformed this space: LLM-powered systems can generate coherent, context-aware dialogue through in-context learning \cite{wu2019interactive}. \textcolor{black}{For example, Purohit et al.\ demonstrated how LLM-based methods can aid word retrieval for individuals with aphasia, suggesting broader applications in reminiscence therapy \cite{10.1145/3584931.3606993}.} Yet, systematic exploration of LLM-powered dialogue for older adults’ reminiscence remains limited. Most current systems neglect the subtle cues—such as attention or engagement—that could help adapt conversation pacing and content in real time \cite{bai2023chatgpt}. This limitation motivates exploring new modalities for implicit interaction and personalization.

\subsection{Eye-Tracking as Alternative Input Modality}

\textcolor{black}{Gaze-driven interaction has long been studied as a means of capturing users’ attentional processes~\cite{yan2024voila,salamatian2025chartgaze}.} More recent work has incorporated eye movements into language-based modeling to reflect user interest in specific image regions and to guide AI attention and prediction\cite{lopez2024seeing}.
Unlike explicit cues such as pointing or verbal references, which require users to externalize well-formed intentions, gaze reflects pre-intentional attention dynamics that emerge during ongoing perception and sense-making~\cite{10.5555/216164.216178,10.1145/3015783.3015794,argyle1976gaze, kleinke1986gaze,Fernandes02102025}.
Lopez-Cardona et al. propose \textit{GazeReward}, which incorporates implicit eye-tracking feedback into an LLM’s reward model, showing that gaze signals significantly improve alignment with human preferences \cite{lopez2024seeing}. Other architectures encode gaze as structured attention dynamics. For example, STARE tokenizes spatio-temporal eye-movement data over predefined regions of interest (ROIs) and feeds these representations into a transformer model to predict user choices \cite{unger2025stare}. Likewise, the \textit{Gaze2AOI} automatically identifies semantically meaningful image regions and links these \textcolor{black}{Areas of Interest} (AOIs) with gaze metrics (e.g. fixation dwell time), illustrating how ROIs can be inferred directly from eye fixations \cite{trajkovska2024gaze2aoi}. 
These works collectively demonstrate that eye movements reflect user interest in specific image regions and that integrating gaze information can guide AI attention and predictions. 

\textcolor{black}{Beyond serving as an alternative input modality, gaze-driven prompts differ fundamentally from explicit cues such as pointing or verbal referencing~\cite{10.1145/3015783.3015794,10.1145/3325285}. While deictic gestures and linguistic references require users to externalize well-formed intentions, gaze operates as an implicit and pre-linguistic signal that reflects emergent attention dynamics during ongoing perception and sensemaking. As such, gaze captures cognitive traces before they are stabilized into explicit intent or symbolic expression, rather than deliberate communicative acts~\cite{argyle1976gaze,kleinke1986gaze,Fernandes02102025}.}

\textcolor{black}{This implicit property makes gaze particularly effective for supporting reminiscence, which relies on revisiting prior attentional and interpretive states rather than recalling fully articulated content. Cognitive studies on episodic memory retrieval show that eye movements often reenact original attention patterns even in the absence of visual stimuli, a phenomenon known as the ``looking-at-nothing'' effect~\cite{tulving1983elements,romaniuk1981looking}. By externalizing attention history, gaze provides access to latent, non-verbal memory cues that are difficult to verbalize yet critical for reflective reconstruction. Consequently, gaze-driven prompts enable AI systems to scaffold reminiscence by aligning with how users previously attended, interpreted, and assigned meaning, rather than solely responding to post-hoc verbal descriptions~\cite{lopez2024seeing, yan2024voila}.}

\section{Expert Interview}\label{Study1}
We conducted expert interviews with four domain experts to explore older adults' nostalgic characteristics and the potential challenges and opportunities they may face when participating in reminiscence activities. Then, we outlined the design considerations and potential challenges of constructing an AI-supported reminiscence system for older adults.

\subsection{Participants and Procedure}
We recruited four experts from local universities and public service agencies for one-to-one, online semi-structured interviews. The panel comprised three professors with relevant research expertise and one social worker with direct practice experience with older adults (see Table~\ref{tab:participants1}).

Interviews began with background questions about each expert’s role and career, followed by prompts on how AI might support old-photo–based digital reminiscence for older adults and the challenges older adults may encounter in this process. Each session lasted approximately 30 minutes (see Appendix~\ref{question}). With permission, all interviews were recorded.

\begin{table}[t]
  \centering
  \caption{Overview of Expert Participants}
  \label{tab:participants1}
  \footnotesize
  \setlength{\tabcolsep}{4pt}
  \renewcommand{\arraystretch}{1.15}

  \begin{tabularx}{\columnwidth}{@{} c c l l X @{}}
    \toprule
    \textbf{ID} & \textbf{G} & \textbf{Role} & \textbf{Affiliation} & \textbf{Expertise} \\
    \midrule
    E1 & F & Professor     & University            & Gerontology; Cognitive psychology \\
    E2 & M & Professor     & University            & Computer science; HCI \\
    E3 & F & Social worker & Public social services & Older-adult counseling; Memory rehabilitation \\
    E4 & M & Professor     & University            & Visual neuroscience \\
    \bottomrule
  \end{tabularx}
\end{table}

\subsection{Data Analysis}

We conducted an inductive thematic analysis~\cite{mcdonald2019reliability} of the expert interviews. Two researchers independently coded the Chinese transcripts and iteratively developed a shared codebook. After calibrating on 20--30\% of the data, we refined coding rules and applied the finalized codebook to the full corpus. Inter-rater agreement on the calibrated subset was substantial (Cohen’s $\kappa$ = .72); disagreements were resolved by consensus and logged. For bilingual reporting, themes and exemplar quotes were translated into English and verified by a second author, with contentious cases resolved via discussion and selective back-translation.

\subsection{Design Considerations}\label{DC}

Based on our findings, we propose two design considerations (DCs) for AI-assisted reminiscence using old photos with older adults. Each DC articulates a rationale and translates it into actionable design rules, supported by participant evidence.

\subsubsection{DC1: Low-Effort, Accessible, and Safe Mixed-Initiative Reminiscence.}
\begin{itemize}
  \item \textbf{Keep interaction natural with minimal cognitive load.} Experts ($E2, E4$) emphasized simplicity and clarity, noting that multi-step, browser-style controls can disrupt the train of thought. The interface should prioritize gaze and speech as the primary modalities and avoid nested menus. As $E2$ put it, \textit{``Some browser-based interfaces pose challenges for older adults and may break their train of thought.''}

  \item \textbf{Provide guided, empathetic AI facilitation.} Experts ($E1, E2$) reported that encouraging prompts may help older adults articulate stories and emotions. The dialogue agent should acknowledge and probe while pacing the conversation (e.g., reflective paraphrasing), and should avoid jargon or rapid-fire questioning. As $E1$ noted, \textit{``Reminiscence aims to evoke personal memories through external memory cues, so the dialogue structure should be inspiring.''}

\item \textbf{Ensure accessibility by default.} Experts ($E2, E4$) highlighted practical accommodations, including high-contrast visuals, large text and targets, clear audio with adjustable volume, and minimal on-screen controls. Where available, a large display can improve legibility and shared viewing ($E2$). Echoing this, $E4$ emphasized that the UI should remain simple and senior-friendly, with reduced visual clutter and clearly distinguishable controls (e.g., larger buttons and straightforward layouts).

\item \textbf{Convey safety and privacy to build trust.} Experts stressed the need to communicate what is recorded and why, and to keep data secure. The system should provide plain-language notices, allow pausing/opt-out at any time, and store logs locally and/or in anonymized form. As $E3$ remarked, \textit{``It is vital to ensure older adults understand the system is secure and will not compromise their privacy.''}
\end{itemize}

\subsubsection{DC2: Content Strategy: Cultural Fit, Effective External Cues, and Chronological Structuring.}
\begin{itemize}
  \item \textbf{Cultural background.} Experts ($E2$, $E3$) recommended selecting photos aligned with local traditions and history so that cues felt familiar and meaningful. As $E3$ noted, \textit{``Selecting old photos that reflect local traditions and history can help them relive and appreciate past experiences and culture.''} $E2$ added, \textit{``Some existing reminiscence tools provide content that local older adults find difficult to relate to.''}

\item \textbf{External memory cues.} Beyond personal albums, era-typical photos can serve as effective cues that stimulate curiosity and enrich storytelling, complementing personal recollection. As $E1$ suggested, era-specific collective photos \textit{``can effectively trigger memories even if older adults have never seen the exact photo before.''} Prior work has also shown that generic old photos used as external cues can support reminiscence outcomes for older adults, including those with dementia~\cite{bender1998therapeutic, photo, creative}.

\item \textbf{Chronological order.} Experts ($E1$, $E3$) recommended organizing selected photos along life-course milestones (e.g., childhood, early adulthood, marriage, parenting, career, retirement) so narratives flowed coherently and context was easier to retrieve ($E3$). Such structuring can reduce cognitive load by providing a predictable narrative scaffold, helping older adults orient themselves in time and smoothly transition between episodes.

\end{itemize}

\begin{figure*}[ht]
\centering
   \includegraphics[height=8cm]{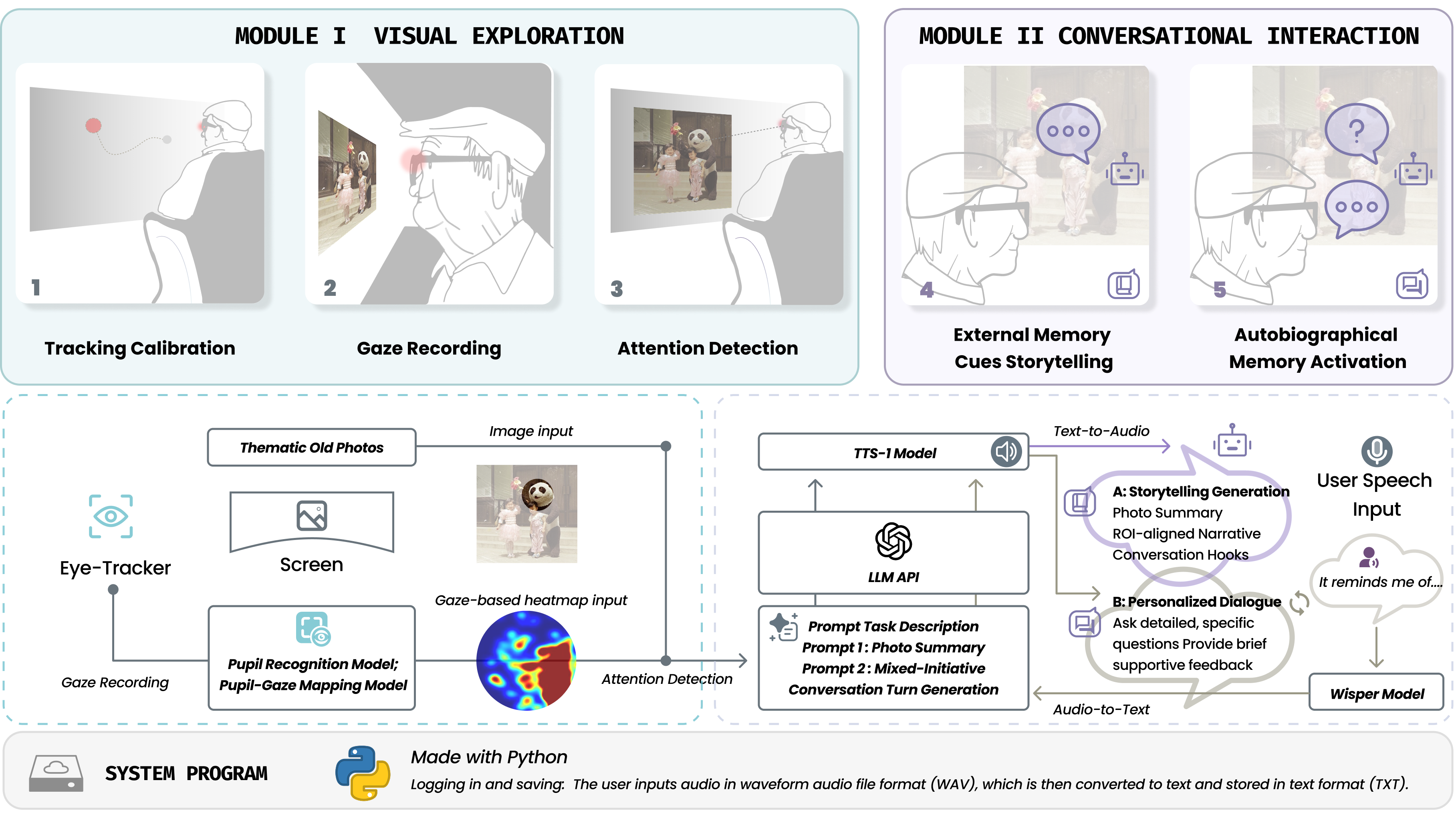} 
\caption{System pipeline of the \textit{Eye2Recall} prototype. The system includes (I) a visual exploration module with eye-tracking calibration, gaze recording, and attention/ROI detection, and (II) a conversational interaction module comprising external-cue--guided storytelling and autobiographical memory activation. Steps~2--5 repeat for each user query in our LLM-powered pipeline.}
\Description{Diagram of the Eye2Recall system pipeline with two modules: a visual exploration module (eye-tracking calibration, gaze recording, attention/ROI detection) and a conversational interaction module (storytelling guided by external cues and autobiographical memory activation). The diagram indicates that the query-handling steps repeat for each user query.}

\label{5} 
\end{figure*}

\section{System Implementation}
Guided by our design considerations (Section~\ref{DC}) from expert interviews, we developed \textit{Eye2Recall}, a prototype AI conversational reminiscence system that supports older adults sharing memories (see Fig.~\ref{5}). The prototype comprises three components: \textbf{\textit{Content Design}}, \textbf{\textit{Gaze-based ROI Detection}}, and \textbf{\textit{Gaze-to-Prompt Adaptation}}.

\begin{figure*}[ht]
\centering
   \includegraphics[height=7cm]{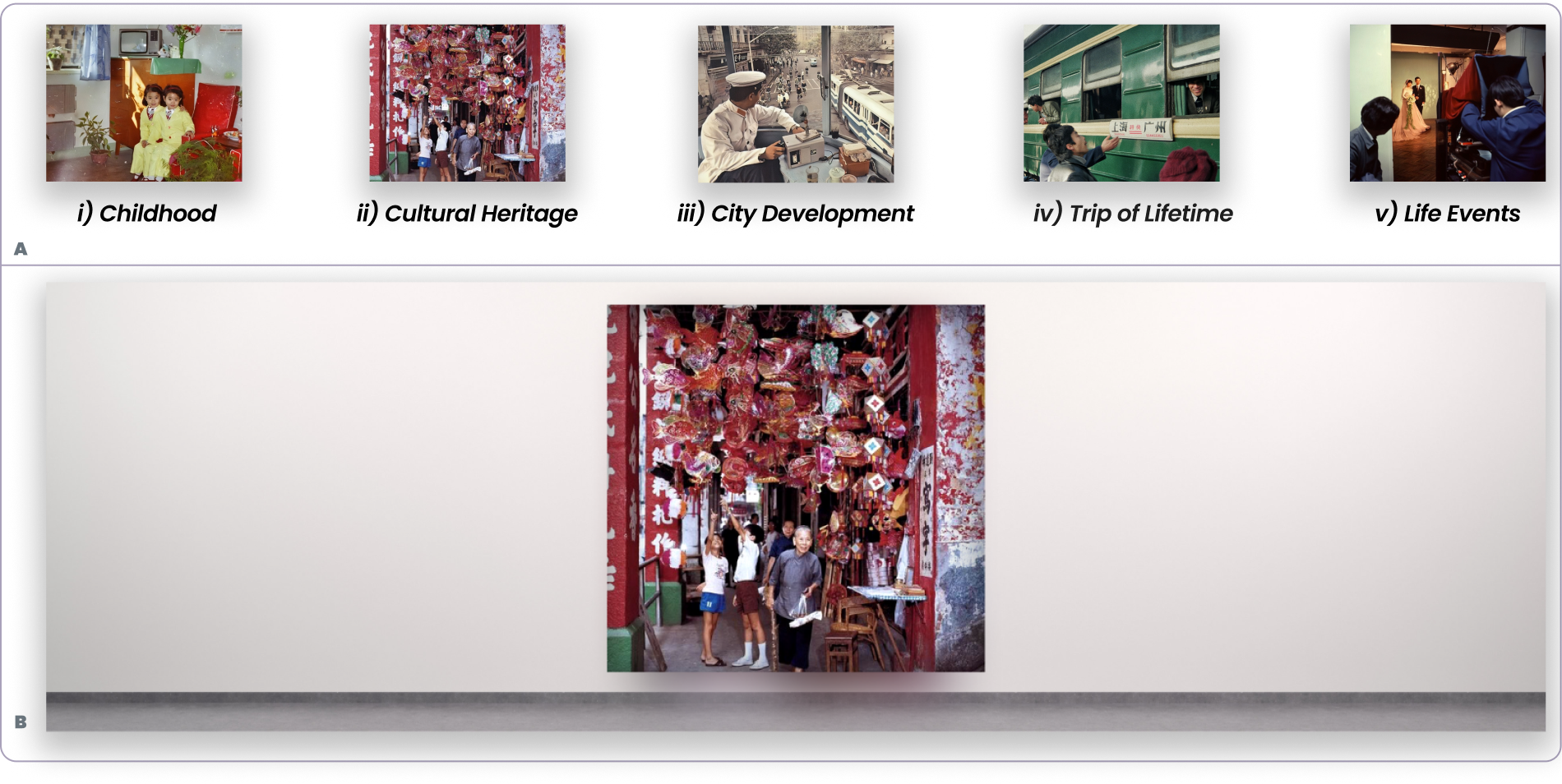}
\caption{(a) Example of visual content categorization, consisting of generic old photos categorized into five themes.
(b) Screenshots of the \textit{Eye2Recall} user interface (UI).}
\Description{(a) Example of visual content categorization, consisting of generic old photos categorized into five themes.
(b) Screenshots of the \textit{Eye2Recall} user interface (UI).}
\label{fig:ui}
\end{figure*}

\subsection{Content Design}\label{visual}
Guided by \textbf{DC2}, we curated the visual content to align with users’ cultural backgrounds, incorporate external memory trigger cues, and present materials in a chronological sequence. In parallel, we designed the user interface to surface these cues clearly and support low-effort, timeline-based navigation.

\subsubsection{Reminiscence Material}
Considering local older adults’ cultural context, we curated approximately 2{,}000 era-typical archival photographs from publicly accessible collections~\cite{NationallibraryofChina, beijingsilvermine, GUANGZHOUDIGITALLIBRARY}. 
The corpus spans the 1970s–1990s and emphasizes everyday life, public events, and locally salient scenes. We applied inclusion criteria (Chinese context; usable resolution; basic metadata such as year or location) and excluded images with sensitive content or clearly identifiable individuals without permission. These photographs serve as external memory cues for photo-elicited reminiscence, providing culturally familiar triggers to support recall and engagement.

To support chronology while retaining meaningful cues, we adopted a two-step organization: (1) theme tagging informed by Boyer’s notion of collective memory in the city and place-based perspectives on memory \cite{boyer1994city,LuShaoming1}; and (2) within each theme, chronological ordering by available metadata (year/period). In line with reminiscence research on external cues \cite{astell2010stimulating}, the era-typical photos were grouped into five themes with operational inclusion criteria:
(i) \textit{Childhood}: early-life settings and relations (home, school, play, caregivers).  
(ii) \textit{Cultural Heritage}: locally salient symbols, festivals, and folk practices (parades, rites, crafts).  
(iii) \textit{Urban Development}: urban change and everyday work contexts (streetscapes, housing, transport, industry).  
(iv) \textit{Migration \& Mobility}: leaving/arriving, travel and relocation for work, study, or life transitions.  
(v) \textit{Life Events}: personal milestones (birthdays, weddings, graduations) depicted in public or family spaces.
Representative examples for each theme are shown in Fig.~\ref{fig:ui}a.

\subsubsection{User Interface}
To ensure accessibility and immersion~\cite{4196176}, we selected photos with clear, salient visual cues (e.g., adequate sharpness and colour, discernible depth and perspective, and coherent spatial layout and motion direction). The user interface displays the historical photo centrally on a large LED screen, while removing non-essential on-screen elements and using a clean, uniform background (see Fig.~\ref{fig:ui}b). This layout keeps the visual focus on the photo content and reduces potential distractions during conversation.

\subsection{Gaze-based ROI Detection}\label{Detection}

Guided by \textbf{DC1}, in the \textit{Visual Exploration Module} (see Fig.~\ref{5}), we use a glasses-based eye tracker to detect users’ visual interests while users view each photo. During viewing, the prototype samples and derives gazes (location and duration) and saccades. Gazes are identified using a duration threshold (e.g., \(\geq\)300\,ms) and clustered on the photo plane to obtain ROIs. From these data, the prototype renders gaze-based heatmaps that summarize the distribution of visual attention (see Appendix~\ref{Calibration Method}). The gaze-based heatmaps are then passed to the \textit{Conversational Interaction Module} to condition the next mixed-initiative turn.

\begin{figure*}[ht]
\centering
\includegraphics[height=8cm]{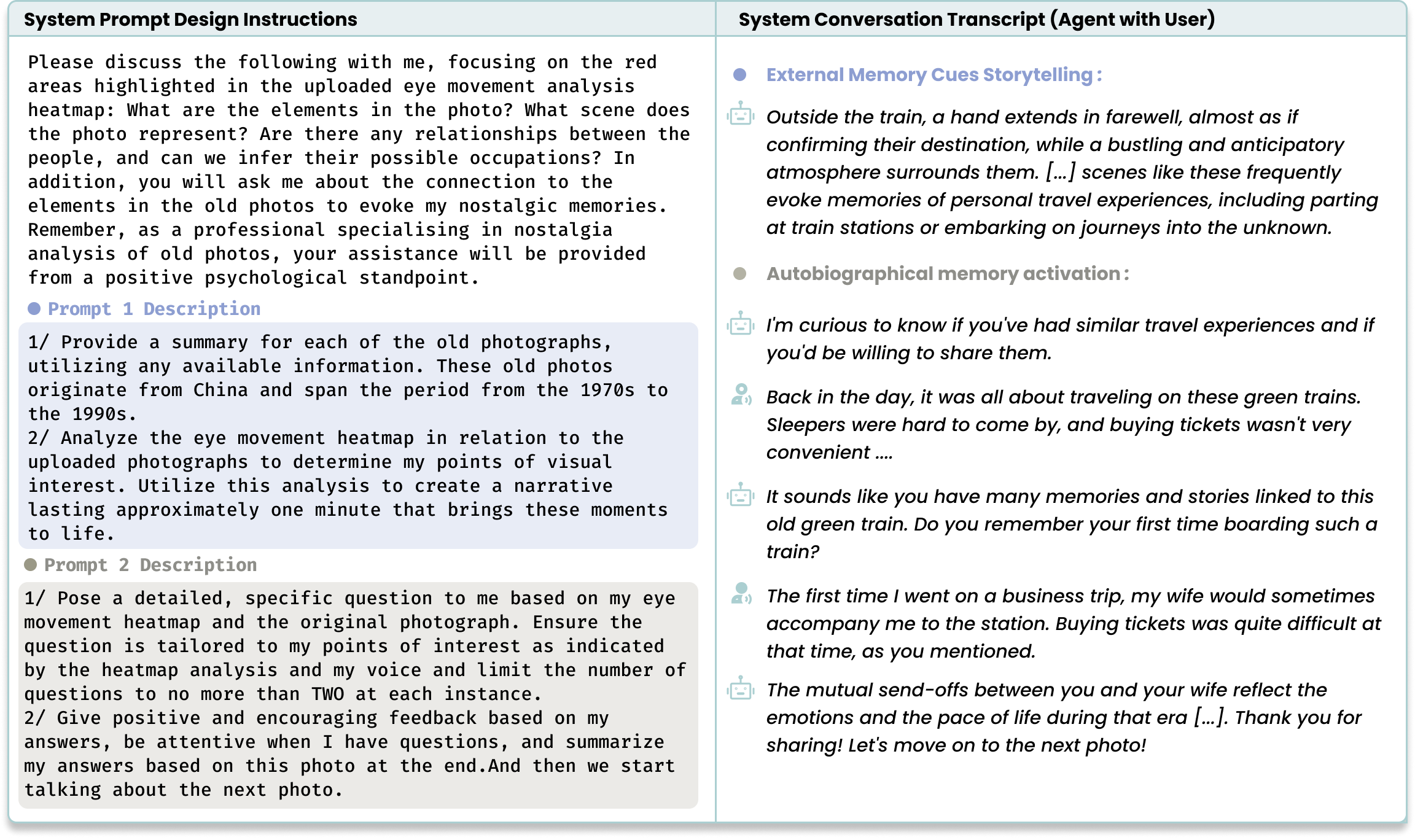} 
\caption{Prompt task instructions and Chat History in \textit{Eye2Recall}.
On the left side, there are descriptions outlining the overall task along with detailed instructions for two distinct modules. On the right side, an example conversation between an LLM-powered agent and a user is provided.} 
\Description{Prompt task instructions and chat history in \textit{Eye2Recall} system.
On the left side, there are descriptions outlining the overall task along with detailed instructions for two distinct modules. On the right side, an example conversation between an LLM-powered agent and a user is provided.}
\label{6} 
\end{figure*}

\subsection{Gaze-to-Prompt Adaptation}\label{Prompt}
We configured the agent with two tailored prompt instructions (see Appendix~\ref{app:prompts} for more details). Leveraging the content of the current photo and the user’s gaze-based heatmap, these prompts drive personalized, context-aware dialogue turns.

\textbf{Prompt~1} provides contextual scaffolding by instructing the agent to \textbf{summarize} the current photo with brief background details to open the conversation (see Fig.~\ref{6}). In pilot tuning, we set the generation parameters to temperature~=~1.0 and a response length of approximately 600 tokens, yielding 1--2\,min spoken summaries and a good balance between breadth and relevance.
  
\textbf{Prompt~2}  conditions the agent on the \textbf{gaze-based heatmap} to ask \textbf{targeted, personalized} questions about the top ROIs, combining external cues with autobiographical recall, hard-capped at two questions per turn (see Fig.~\ref{6}). Parameters were set to temperature~=~0.5 and approximately 200 tokens, yielding concise 20--40\,s follow-ups that maintained attention without overloading users.

Text outputs were synthesized to speech with TTS-1 and presented to the user. Each photo was handled in two dialogue turns (Prompt~1 followed by Prompt~2), after which the system automatically advanced to the next photo. Conversation transcripts between the agent and the participant are shown in Fig.~\ref{6}.

\subsection{Implementation Details}\label{Apparatus1}
The hardware setup includes a glasses-based eye tracker, a large LED display, external stereo speakers, a wireless lapel microphone, and a host workstation for synchronized data capture. The software stack uses a vision-capable LLM API\footnote{GPT-4-vision-preview API Documentation. Available at: \url{https://platform.openai.com/docs/guides/vision/vision}} for image understanding and prompt generation, a text-to-speech model~\footnote{TTS-1 API Documentation. Available at:~\url{https://platform.openai.com/docs/models/tts-1}} for spoken output, and an automatic speech recognition model~\footnote{Whisper API Documentation. Available at:~\url{https://platform.openai.com/docs/guides/speech-to-text}} for transcribing participants' speech. Detailed eye-tracker specifications and calibration procedures are provided in Appendix~\ref{Device}. All data were de-identified and stored on an encrypted local drive; personally identifying information was removed after use. API calls were configured for transient processing only, and local session logs can be purged at the end of each study session.

\section{User Study}
We conducted a user study (N=12) to evaluate the \textit{Eye2Recall}, focusing on usability, workload, engagement, and affective outcomes. The study design and procedure are summarized in Fig.~\ref{4}. The study protocol was reviewed and approved by the institutional ethics committee of HKUST (Guangzhou) (Approval No. HKUST(GZ)-HSP-2025-0394), and all participants provided informed consent and could withdraw at any time without penalty. We collected only study-relevant data (questionnaires, interaction logs, and interviews); personally identifiable information was not stored with the research data. All datasets were pseudonymized using participant codes, stored on encrypted drives with access restricted to the research team, and reported only in aggregate form or through anonymized quotations. A qualified social worker was available on call to support participants and to implement a predefined distress-and-pausing procedure when needed.

\subsection{Participants}

We recruited a sample of 12 participants (5 male and 7 female) through online social media recruitment (see Table~\ref{table_demographics}). Participants were aged between 60 and 84 years ($M = 70.42, SD = 6.08$). 
Participants reported normal or corrected-to-normal vision and hearing, and had no motor impairments that would prevent operation of the study apparatus. 
Each participant received 200 RMB (about 27 USD) as compensation.

According to the Chinese version of the \textit{Mini–Mental State Examination (MMSE)}~\cite{MMSE} and our screening protocol, scores of 28–30 were treated as within the normal range, whereas scores $\leq 27$ suggested a risk of mild cognitive impairment (MCI); thresholds may vary by age and education, and the MMSE is a screening tool rather than a diagnostic test. Across all participants ($M = 28.92$, $SD = 1.08$), 10 screened within the normal range and 2 screened at risk of MCI. No participant scored $< 21$ under our protocol; thus, no case met our criterion for severe impairment risk.

\begin{table}[t]
\centering
\caption{Demographics of Older Adult Participants.}
\label{table_demographics}
\footnotesize
\setlength{\tabcolsep}{3pt}
\renewcommand{\arraystretch}{1.1}

\begin{tabularx}{\columnwidth}{@{}>{\centering\arraybackslash}p{0.8cm}
                                >{\centering\arraybackslash}p{0.7cm}
                                >{\centering\arraybackslash}p{0.8cm}
                                >{\raggedright\arraybackslash}X
                                >{\centering\arraybackslash}p{1.1cm}@{}}
\toprule
\textbf{ID} & \textbf{Gender} & \textbf{Age} & \textbf{Education Level} & \textbf{MMSE} \\
\midrule
P1  & M & 73 & Teacher's training college & 30 \\
P2  & F & 73 & Teacher's training college & 29 \\
P3  & F & 64 & High school                & 29 \\
P4  & M & 84 & High school                & 27* \\
P5  & F & 67 & High school                & 30 \\
P6  & M & 70 & Teacher's training college & 29 \\
P7  & F & 70 & University                 & 29 \\
P8  & F & 60 & High school                & 27* \\
P9  & M & 76 & High school                & 30 \\
P10 & M & 67 & University                 & 30 \\
P11 & F & 69 & University                 & 28 \\
P12 & F & 72 & High school                & 29 \\
\bottomrule
\end{tabularx}

\vspace{2pt}
\begin{minipage}{\columnwidth}
\footnotesize \textit{Note.} *MMSE scores for P4 and P8 are below the normal range, indicating potential risk of mild cognitive impairment (MCI).
\end{minipage}
\end{table}

\subsection{Evaluation Dimensions}

\subsubsection{Quantitative Evaluation}
We adopted three key dimensions for quantitative evaluation: Perceived usability, Affect and Task Performance. 

The \textbf{Perceived usability} dimension captured participants’ subjective experiences of usability and task load. To measure this, we used a custom user experience questionnaire and the \textit{NASA Task Load Index} (NASA-TLX)~\cite{hart2006nasa}, both rated on a 7-point Likert scale. The custom questionnaire assessed \textit{accessibility, effectiveness, and overall user experience} (see Appendix~\ref{Questionnaire}), adapted from the \textit{System Usability Scale} (SUS)~\cite{brooke1996sus}. 

The \textbf{Affect} dimension captured participants’ emotional state prior and post to system use, measured with the \textit{Positive and Negative Affect Schedule} (PANAS); PA and NA subscales~\cite{panas}. This scale, comprising 20 items, equally measures positive affect (e.g., \textit{excited, inspired}) and negative affect (e.g., \textit{upset, anxious}) and is well-regarded for capturing emotional nuances~\cite{schmukle2002relationship}

\textbf{Interaction dynamics.}
We analyzed eye-tracking data for all participants, including gaze-based heatmaps, gaze ratios, saccade frequencies, response duration and response latency. Two researchers examined gaze distributions to identify common and divergent gaze patterns across participants.
We applied a descriptive, mixed-methods approach to the conversational content. We used the text-mining platform \textit{Weiciyun}\footnote{\textit{Weiciyun},  Available at~\url{https://www.weiciyun.com}.} to conduct word-frequency~\cite{baron2009word,medhat2014sentiment}.

\subsubsection{Qualitative Evaluation}
To complement these quantitative methods, we conducted a \textbf{semi-structured interview} to gather qualitative feedback on participants’ experiences, providing deeper insights into strengths, weaknesses, and improvement areas. All recordings were transcribed using a commercially available automatic speech recognition (ASR) system~\footnote{\textit{iFLYTEK}. Available at~\url{https://www.iflyrec.com/zhuanwenzi.html}}.
Two researchers performed thematic analysis~\cite{mcdonald2019reliability} with an iterative codebook. Disagreements were resolved through discussion until consensus. The qualitative results were used to explain and extend the quantitative findings, with particular attention to concrete moments where gaze-driven prompts supported reminiscence or caused friction.

\begin{figure*}[h]
\centering
\includegraphics[width=\textwidth]{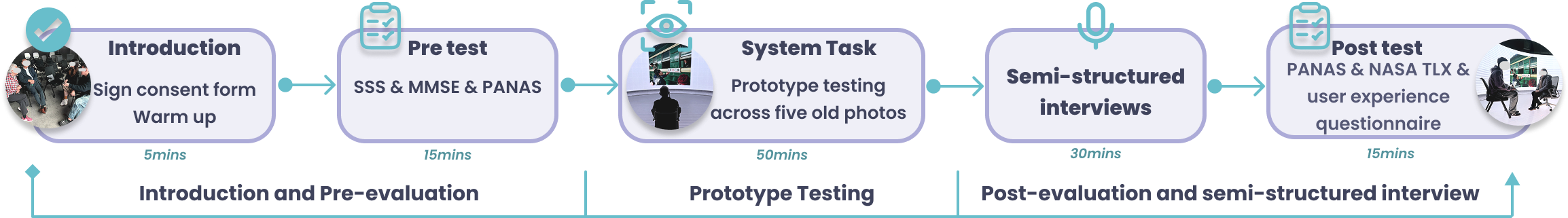} 
\caption{The user study process comprised three phases: (1) the introduction and pre-evaluation, (2) the prototype testing, and (3) semi-structured interviews with the post-evaluation. Each participant participated for approximately 1.5 hours.} 
\Description{The user study process comprised three phases: (1) the introduction and pre-evaluation, (2) the prototype testing, and (3) semi-structured interviews with the post-evaluation. Each participant participated for approximately 1.5 hours.} 
\label{4} 
\end{figure*}

\subsection{Procedure}
The study was conducted by two researchers, with a professional social worker available on call, in a university laboratory. Each session lasted approximately 1.5 hours and followed three phases (see Fig.~\ref{4}).

\subsubsection{Introduction and pre-evaluation}
We introduced the system and procedures, obtained written informed consent, and administered a demographics form. The Chinese version of MMSE was used to screen cognitive status. At the beginning of the session, we assessed participants’ alertness using the \textit{Stanford Sleepiness Scale} (SSS)~\cite{shahid2012stanford}. Pre-session affect was measured using the PANAS. To reduce burden, all items were presented in large fonts with high contrast. The researcher could read items aloud on request.

\subsubsection{Prototype testing}
The test phase lasted approximately 50 minutes and followed a two-stage workflow—\textit{Visual Exploration} and \textit{Conversational Interaction}—as summarized below.

\begin{figure*}[h]
\centering
\includegraphics[width=\textwidth]{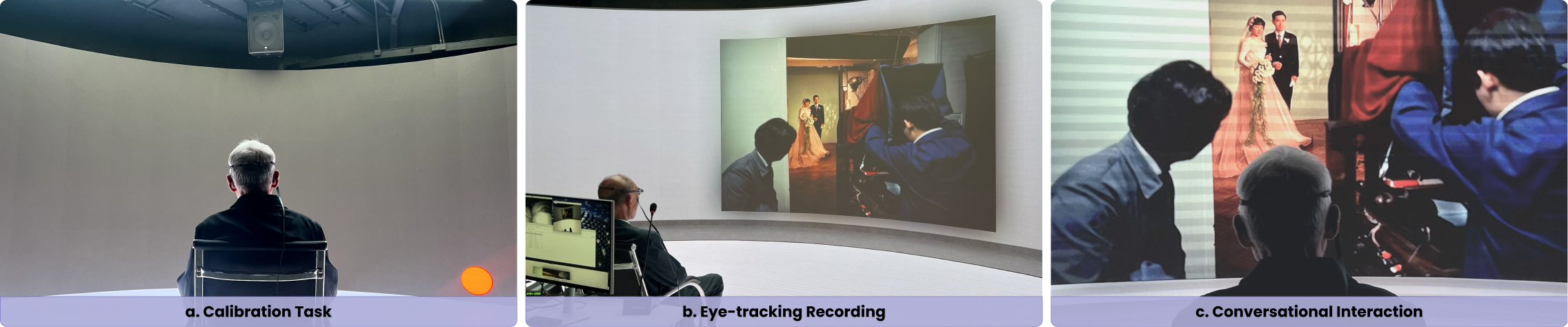} 
\caption{In the prototype testing, we selected three representative photos of the live interaction scenarios: (a) Calibration task, (b) Eye tracking recording, and (c) Conversational interaction.} 
\Description{In the System Testing, we selected three representative photos of the live interaction scenarios: (a) Calibration task, (b) Eye tracking recording, and (c) Conversational interaction.} 
\label{2} 
\end{figure*}

\begin{itemize}
  \item \textbf{\textit{Visual Exploration}} :
  (i) Participants were seated comfortably, fitted with the glasses-based eye tracker, and completed a standard calibration. They then rested for about one minute with eyes closed (see Fig.~\ref{2}(a)).
  (ii) Participants viewed five era-typical photos on the LED display, 1 minute per photo, in a predetermined order while an experimenter supervised the procedure (see Fig.~\ref{2}(b)).

  \item \textbf{\textit{Conversational Interaction}}: 
(i) The agent delivered a brief spoken summary (\textit{Prompt~1}) of the current photo (approximately 1–2 minutes) via external speakers (see Fig.~\ref{2}(c)). 
(ii) Using gaze-based heatmap, the agent then asked up to two targeted questions about each photo (\textit{Prompt~2}), and participants responded via a wireless microphone. After two dialogue turns per photo, the system automatically advanced to the next photo.
\end{itemize}

\subsubsection{Post-evaluation and semi-structured interview}
After the test phase, participants completed the custom user experience questionnaire, PANAS and NASA–TLX. A semi-structured interview (approximately 30 minutes) was then conducted to elicit perceptions of usability, workload, pacing, prompt relevance, and suggestions for improvement.

\section{Findings}

This user study aimed to evaluate the older adult component of \textit{Eye2Recall}. Overall, the results indicate that the pilot system demonstrated good usability (Section~\ref{111}), enabled older adults to actively engage in LLM-mediated conversations through gaze-based interaction (Section~\ref{FD}), and provided perceived emotional and well-being support (Section~\ref{FD2}).

\subsection{Impact on Perceived Usability}\label{111}


\begin{figure*}[h]
\centering
\begin{minipage}{0.48\textwidth}
  \centering
  \includegraphics[width=\linewidth]{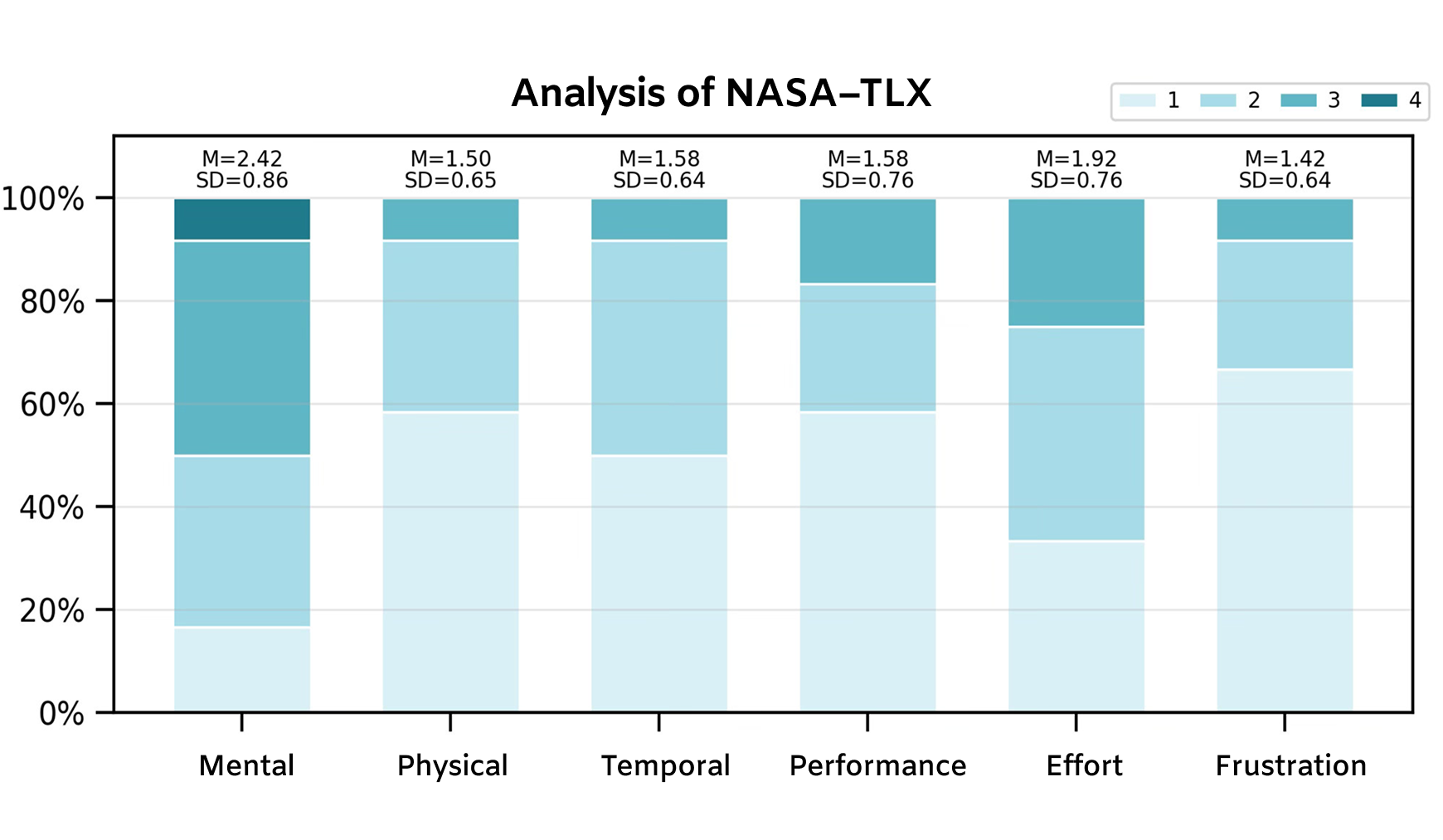}
\captionof{figure}{Stacked bar chart of NASA--TLX ratings across six workload dimensions (7-point Likert scale; 1=best, 7=worst; $n=12$). No responses fell in the high range (scores 5--7).}
  \Description{Stacked bar chart of NASA--TLX ratings across six workload dimensions (7-point Likert scale; 1=best, 7=worst; $n=12$). No responses fell in the high range (scores 5--7).}
  \label{fig:nasa}
\end{minipage}
\hfill
\begin{minipage}{0.48\textwidth}
  \centering
  \includegraphics[width=\linewidth]{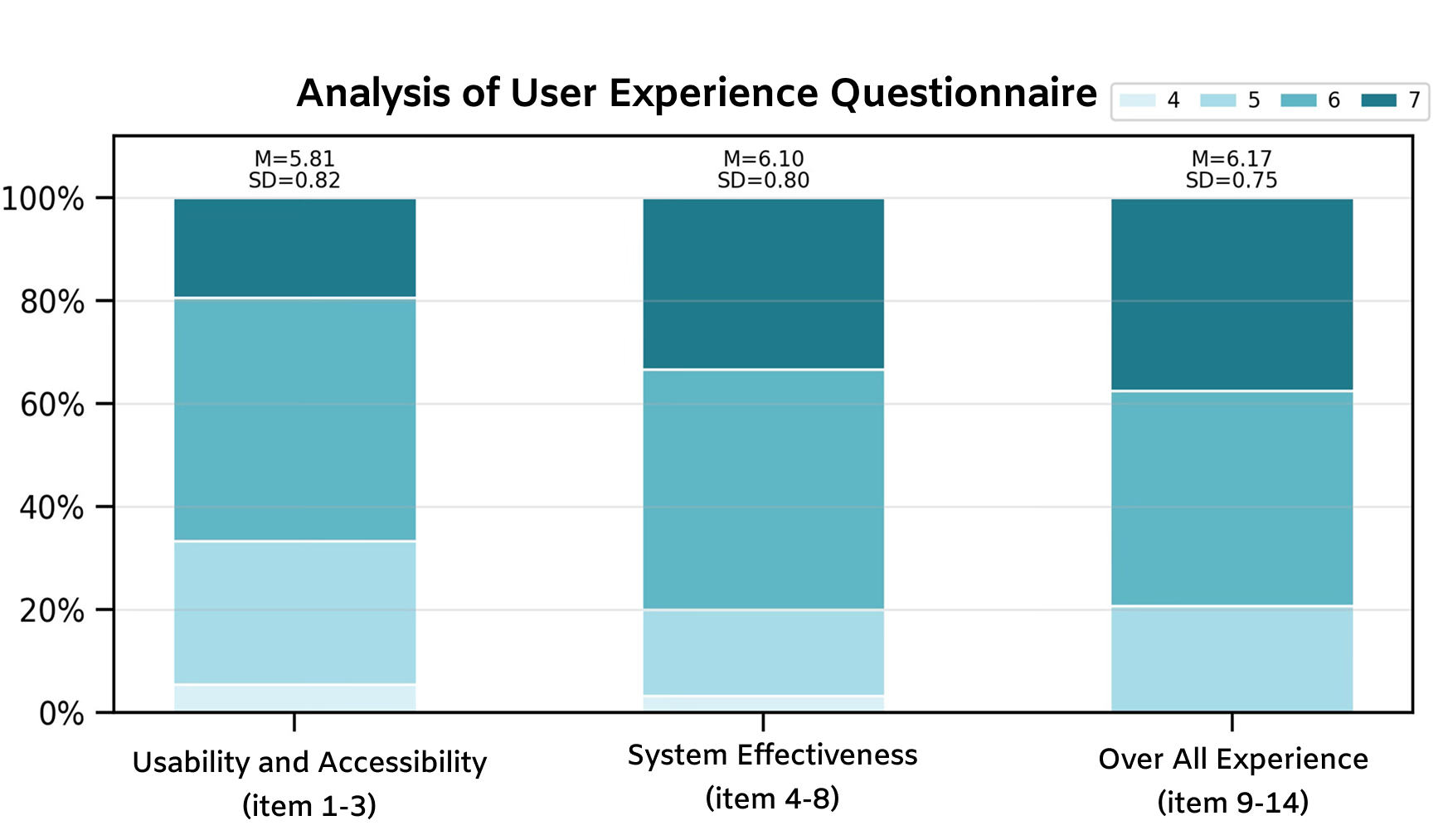}
  \captionof{figure}{Stacked bar chart of the Custom User Experience Questionnaire ratings (7-point Likert scale; 1=strongly disagree, 7=strongly agree; $n=12$). No ratings fell in the low range (scores 1--3).
}
  \Description{Stacked bar chart of the Custom User Experience Questionnaire ratings (7-point Likert scale; 1=strongly disagree, 7=strongly agree; $n=12$). No ratings fell in the low range (scores 1--3).}
  \label{UEQ}
\end{minipage}

\end{figure*}

\subsubsection{NASA-TLX Questionnaire}
We measured participants’ perceived \textbf{workload} using the NASA--TLX. 
Overall workload was low (\(M=1.74\), \(SD=0.28\)), all six sub-scale means were well below the midpoint (see Fig.~\ref{fig:nasa}), indicating consistently low perceived workload.
By subscale, \textit{Mental Demand} had the highest mean (\(M=2.42\), \(SD=0.86\)), which suggest that participants invested cognitive effort in interpreting photos and formulating narratives. It was followed by \textit{Effort} (\(M=1.92\), \(SD=0.76\)). \textit{Temporal Demand} (\(M=1.58\), \(SD=0.64\)) and \textit{Performance} (\(M=1.58\), \(SD=0.76\)) were lower, while \textit{Physical Demand} (\(M=1.50\), \(SD=0.65\)) and \textit{Frustration} (\(M=1.42\), \(SD=0.64\)) were the lowest. 

Participants’ interview feedback aligned with the NASA-TLX results, indicating generally low perceived workload during the photo-based conversations. Beyond cognitive workload, participants also commented on interaction comfort and modality quality. Most participants described the glasses-based eye tracker as comfortable (e.g., $P7$ \textit{“felt like a regular pair of glasses,”} and $P3$ noted, \textit{“most of the time I didn’t even notice it.”}). However, two participants ($P10$, $P12$) anticipated discomfort with prolonged wear; as $P10$ noted, \textit{“I might lose patience if I wore it for a long time.”} 

\subsubsection{Custom User Experience Questionnaire}

To assess participants’ \textbf{user experience}, we asked them to rate the system on a 7-point questionnaire covering three dimensions (14 items in total). 
Across dimensions, \textit{Overall User Experience} received the highest rating ($M=6.17$, $SD=0.75$), followed by \textit{System Effectiveness} ($M=6.10$, $SD=0.80$) and \textit{Usability and Accessibility} ($M=5.81$, $SD=0.82$). 
Shapiro--Wilk tests on participant-level dimension means indicated no evidence of non-normality for the Custom User Experience Questionnaire dimensions (all $p>.28$). One-sample $t$-tests against the scale midpoint (4) further showed that all three dimensions were significantly above neutral: Usability and Accessibility ($t(11)=8.90$, $p<.001$), System Effectiveness ($t(11)=11.25$, $p<.001$), and User Experience ($t(11)=12.62$, $p<.001$). These results suggest that participants perceived the system as usable, effective, and engaging during the reminiscence activities.

\begin{figure*}[h]
\centering
\includegraphics[width=\textwidth]{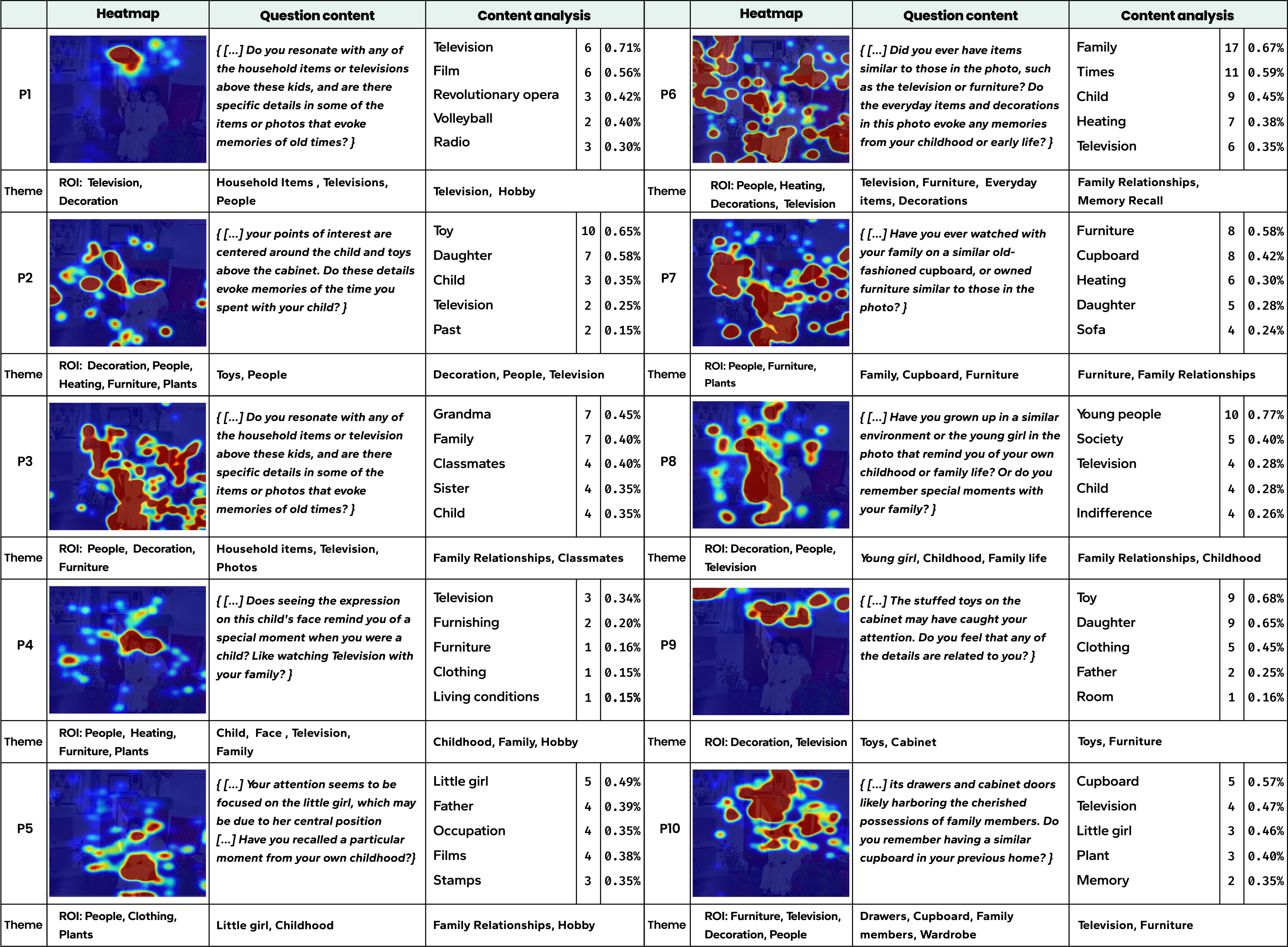} 
\caption{This figure illustrates example gaze-based heatmaps from 10 participants (out of 12 in total) while viewing \textit{Childhood}-themed photos, together with an analysis of their conversational content. The analysis presents the agent’s first-round prompt questions, the frequency counts of the five most common nouns in participants’ dialogues with the AI agent, and their corresponding TF–IDF (term frequency–inverse document frequency) values.}
\Description{This figure illustrates example gaze-based heatmaps from ten participants (out of twelve in total) while viewing \textit{Childhood}-themed photos, together with an analysis of their conversational content. The analysis presents the agent’s first-round prompt questions, the frequency counts of the five most common nouns in participants’ dialogues with the AI agent, and their corresponding TF–IDF (term frequency–inverse document frequency) values.}
\label{zhuanlu}
\end{figure*}

\subsection{Analysis of Task Performance}\label{FD}
During the study, all participants completed the five photo-based conversation blocks. In Section~\ref{sec:gazeinput}, we summarize how gaze input related to the conversational themes that participants reported. Section~\ref{6.2.2} then examines block-wise interaction dynamics using objective measures, including gaze-based engagement and conversational timing.

\subsubsection{Gaze-informed prompts align with conversational themes and enable more tailored dialogue}~\label{sec:gazeinput}
To examine how gaze input relates to conversational themes, we triangulated participants’ gaze heatmaps with transcripts of their dialogues with the agent (Fig.~\ref{zhuanlu}). For each thematic photo, we first identified a set of salient content themes (e.g., objects/people/places visible in the photo) and then mapped gaze regions of interest (ROIs) to these themes based on the referenced visual elements. This allowed us to compare which photo elements attracted visual attention and which elements became the focus of subsequent conversation. 

In the \textit{Childhood} photo, we identified six dominant themes: \textit{People, Furniture, Television, Plants, Heating}, and \textit{Decoration}. Participants whose gaze was concentrated on a small subset of ROIs often produced dialogue that was correspondingly object-specific. For instance, $P1$ primarily attended to \textit{Television} and \textit{Decoration}, and the ensuing dialogue frequently referenced household items and television-related memories (e.g., \textit{television} and \textit{film} were each mentioned six times in the transcript). Similar photo-element-to-dialogue alignment was observed for $P3, P4, P5, P9$, and $P10$, suggesting that gaze cues can help prioritize content for more tailored follow-up questions.

In contrast, when participants’ ROIs were scattered across multiple objects, the system generated more general and less personalized questions. For example, $P6$’s ROIs covered nearly all thematic elements, and the agent’s questions were correspondingly more general. $P6$’s responses contained frequently mentioned but semantically broad words such as \textit{Family} (17 times), \textit{Times} (11 times), and \textit{Child} (9 times). Their low TF–IDF values further indicated less object-specific language.

\textbf{Participants' Perspectives on Gaze-Driven Personalized Conversations.}
Most participants expressed satisfaction with the quality of the gaze-driven personalized conversations facilitated by the LLM agent. This was also reflected in  \textit{\textbf{Q4}} (\textit{The system accurately understands what I say}, $M=6.00$, $SD=0.95$) and \textit{\textbf{Q6}} (\textit{The system detects my visual interests and directs the conversation accordingly}, $M=5.58$, $SD=0.76$). Participants noted that the prototype often steered the dialogue toward content they found personally meaningful. For example, $P1$ explained:

\begin{quote}
\textit{"I feel like it truly understands what interests me. It (Agent) starts by asking about TV shows. The old TV brings back so many memories, which makes me keep looking at it."}
\end{quote}

Similarly, $P4$ reported that the agent helped surface overlooked details and sustain engagement:

\begin{quote}
\textit{"When I used to look at old photos, I often overlooked the details from the past. However, the AI brought up topics that genuinely intrigued me, making me feel more engaged and connected."}
\end{quote}

However, two participants ($P2$ and $P5$) raised concerns about whether some questions reliably reflected their gaze. As $P5$ noted, \textit{"I am not sure if all the questions were based on what I saw."} Overall, these responses suggest that gaze-conditioned prompting can make reminiscence conversations feel more personalized, while highlighting the need for more transparent and robust gaze-to-question grounding to build user trust.

\begin{figure*}[h]
\centering
\includegraphics[width=\textwidth]{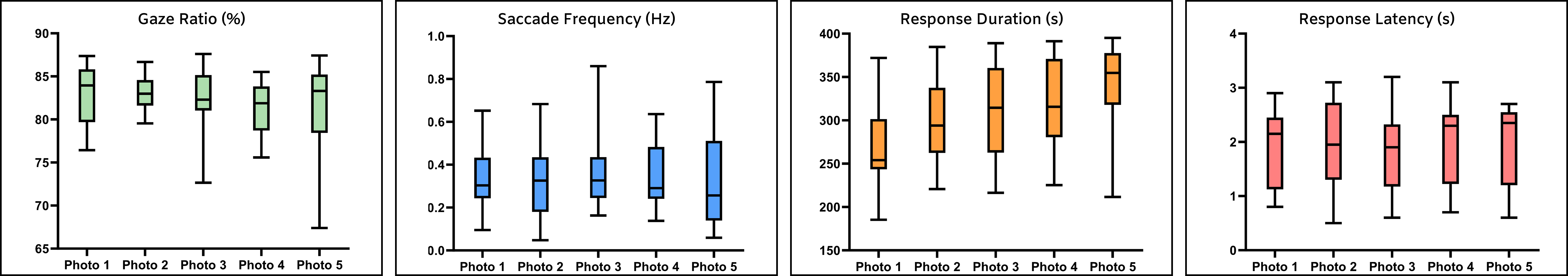} 
\caption{Sequential block effects across five photo blocks (Photo 1--5). Boxplots summarize gaze ratio (\%), saccade frequency (Hz), response duration (s), and response latency (s) for $n=12$.}
\Description{Sequential block effects across five photo blocks (Photo 1--5). Boxplots summarize gaze ratio (\%), saccade frequency (Hz), response duration (s), and response latency (s) for $n=12$.}
\label{8}
\end{figure*}

\subsubsection{Analysis of Gaze Engagement and Conversational Behavior}\label{6.2.2}

To assess whether gaze engagement and conversational timing showed systematic changes across the five sequential photo blocks (potential order or time-on-task effects), we analyzed gaze-based engagement (gaze ratio, saccade frequency) and conversational timing (response duration, response latency) across Photo~1--Photo~5 (see Fig.~\ref{8}); operational definitions and computation details for these metrics are provided in Appendix~\ref{app:metrics}. Given the repeated-measures design across Photo~1--Photo~5 and the small sample size, we used nonparametric Friedman tests~\cite{zimmerman1993relative} to assess block-wise differences.
 
Overall, gaze-based engagement remained stable across Photo~1--Photo~5. \textbf{Gaze ratio} was consistently high, ranging from 81.14\% to 82.99\% (a 1.85 percentage-point span). Mean gaze ratio showed only a slight decrease from Photo~1 (82.81\%) to Photo~5 (81.48\%). Variability was highest in Photo~5 ($SD=5.28$), indicating greater between-participant dispersion in that block rather than a clear monotonic decline over time. Consistent with these descriptive patterns, the Friedman test indicated no significant block effect on gaze ratio, $\chi^2(4)=6.73$, $p=.151$, Kendall's $W=.14$.

\textbf{Saccade frequency} was likewise stable across Photo~1--Photo~5 (0.32--0.38\,Hz), showing no progressive decrease from Photo~1 to Photo~5 (0.33 to 0.34\,Hz). A small descriptive uptick in the middle blocks (Photo~3--4) was observed, but this pattern was not reliable. The Friedman test indicated no significant block effect on \textbf{saccade frequency}, $\chi^2(4)=2.07$, $p=.723$, Kendall’s $W=.04$.

In contrast, \textbf{response duration} increased across the session, from Photo~1 ($M=262.98$\,s, $SD=49.24$) to Photo~5 ($M=341.82$\,s, $SD=51.12$), a net increase of 78.84\,s. This increase may reflect participants becoming more comfortable with the activity over time and/or later photos eliciting more elaborate narratives; however, we cannot fully disentangle order effects from photo-specific content in the present design. The Friedman test indicated a significant block effect on response duration, $\chi^2(4)=25.00$, $p<.001$, Kendall’s $W=.52$.
\textbf{Response latency} remained stable across Photo~1--Photo~5 (means 1.85--2.05\,s), with only a small change from Photo~1 to Photo~5 (1.88 to 2.02\,s), suggesting no clear slowing in response initiation over time. The Friedman test was not significant for \textbf{response latency}, $\chi^2(4)=3.96$, $p=.411$, Kendall’s $W=.08$.

\textbf{Participants also reported a gradual increase in their willingness to express themselves and sustain attention over the five photo blocks.} For example, $P9$ appreciated the chronological sequencing from past to present, describing it as ``like screening a series of old photos,'' and noted that gaze-based interaction felt effortless: 
\begin{quote}
\textit{``I was satisfied with the ordering of the five photos from past to present. The overall experience was pretty good. Especially, this gaze-based form felt easy for me, because while I was watching, it (Agent) already knew what I was paying attention to''} ($P9$). 
\end{quote}

In addition, some participants desired greater agency over the conversational flow ($P1, P7, P9$). As $P7$ remarked, \textit{``I feel that two rounds of dialogue for each photo may not be enough for me''}. These reflections suggest that future versions could offer more user control over parameters such as the number of dialogue turns per photo and the time spent on each image, allowing the system to better accommodate individual pacing and narrative depth.

\subsubsection{Linking log-derived metrics to Subjective Outcomes}
We explored whether interaction logs related to subjective outcomes by correlating participant-level mean log metrics (gaze ratio, saccade frequency, response latency, response duration) with overall NASA-TLX workload and overall user experience (UX) ($n=12$). Spearman rank correlations were used and Holm correction~\cite{abdi2010holm} was applied within each set of four tests. No associations remained significant after correction (NASA: all $p_{\mathrm{Holm}} \ge .380$; UX: all $p_{\mathrm{Holm}} \ge .492$). The strongest uncorrected effects were response latency with NASA ($\rho=-.504$, $p=.095$, $p_{\mathrm{Holm}}=.380$) and gaze ratio with UX ($\rho=-.470$, $p=.123$, $p_{\mathrm{Holm}}=.492$).

\subsection{Emotional and Reflective Benefits}\label{FD2}

\subsubsection{Enhancing emotional well-being through positive memory recall} \label{Well-being:}

\begin{figure}[t]
  \centering
  \includegraphics[width=\columnwidth]{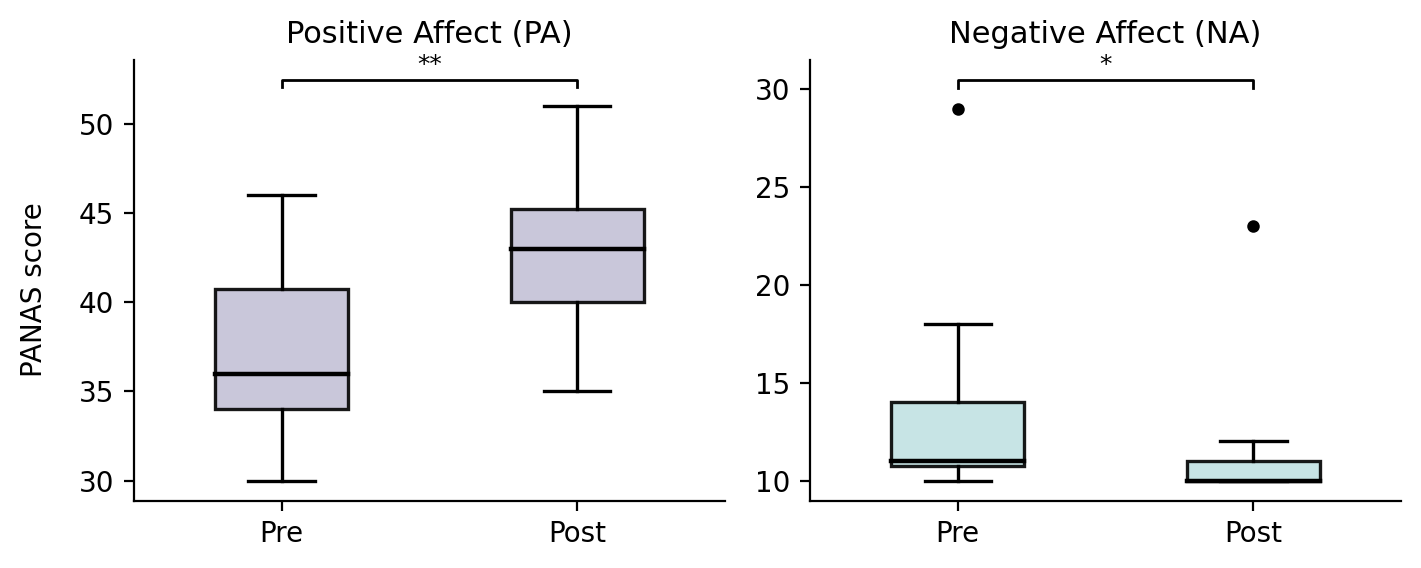}
  \caption{PANAS pre--post score results ($n=12$): Positive affect (PA) increased and negative affect (NA) decreased. Statistical significance is indicated as *$p<.05$, **$p<.01$.}
  \Description{PANAS pre--post results ($n=12$): Positive affect (PA) increased and negative affect (NA) decreased. Statistical significance is indicated as *$p<.05$, **$p<.01$.}
  \label{fig:system2}
\end{figure}

To examine whether interacting with \textit{Eye2Recall} was associated with short-term affective changes, we assessed participants’ pre--post affect using PANAS. Overall, participants reported higher positive affect and lower negative affect immediately after the brief session. We assessed affect pre--post using PANAS (see Fig.~\ref{fig:system2}).
\textbf{Positive Affect (PA)} increased from $M_{\text{pre}}=37.42$ ($SD=5.25$) to $M_{\text{post}}=42.75$ ($SD=4.39$); a paired-samples $t$-test indicated a significant improvement, $t(11)=3.17$, $p=.009$, with a large within-subject effect size ($d_z=0.92$). 
\textbf{Negative Affect (NA)} decreased from $M_{\text{pre}}=13.42$ ($SD=5.43$) to $M_{\text{post}}=11.50$ ($SD=3.68$). 
Given the small sample size, we additionally ran Wilcoxon signed-rank tests as a robustness check; results were directionally consistent (PA: $p=.012$; NA: $p=.022$).

Participants described their experience with positive descriptions such as \textit{"cozy," "warm,"} ($P2, P6$) and \textit{"touching,"} ($P4, P9$), likening it to \textit{"a dialogue between time and space"} ($P5$). Participants felt that the system provides meaningful memories and emotional experiences (\textit{\textbf{Q7}, The system provides meaningful memories and emotional experiences}.  $M=6.17$, $SD=0.58$). After using the system, they felt happy, and their psychological well-being was enhanced (\textit{\textbf{Q8}, After using the system, I feel happy and my psychological well-being is enhanced}. $M=6.17$, $SD=0.80$). $P4$ added \textit{"it gave me much encouragement and was willing to listen to my story, it makes me feel happy".} $P6$ commented that:
\begin{quote}
\textit{"In everyday life, my children work in other cities, so I rarely have someone who truly listens to me. It felt wonderful to have someone willing to hear me out. When I told the story of how my mother struggled to support our whole family when she was young, the agent gave her heartfelt recognition, and I was deeply moved."}
\end{quote}

\subsubsection{Deepening reflection through historical context provided by the LLM agent}

By presenting brief, photo-linked historical prompts and then asking specific questions, the system encouraged reflective remembrance, helping participants to connect personal experiences to wider social change, organise their stories and express their meanings rather than simply recalling facts. Participants shifted their focus to broader social changes, such as household technologies, festive practices and neighbourhood life, when interacting with the LLM agent. Participants connected personal episodes to broader social change and articulated meanings beyond factual recall. P5 called these reflections 
\textit{“priceless treasures,”} while P6 said: 
\begin{quote}
\textit{“After talking with the agent, I rediscovered the atmosphere of Spring Festival in my childhood, even though life was difficult at that time,”} 
\end{quote}
noting that brief introduction about regional folk customs felt \textit{“intimate.”} $P11$ similarly remarked that the experience made previously overlooked parts of the past more salient and memorable.
\subsection{Suggestions for System Improvement}

\subsubsection{Enhancing emotional responsiveness in reminiscence dialogue}\label{cry}

The agent’s encouraging style during dialogue appeared to support affective engagement and reflection. Several participants ($P1, P6, P8, P11$) suggested that improving emotion detection and response would make the experience more engaging. As $P11$ put it, 
\begin{quote}
\textit{"I hoped the agent could vary its prosody and offer more personalized reactions based on user feedback; for example, when I recounting a wedding story, the agent’s tone should sound warm and celebratory to match the moment."}
\end{quote}

Two participants ($P6$ and $P8$) became tearful and choked up during the conversational phase. $P8$ described the experience as immersive and emotionally real, explaining that \textit{``I cried mainly because I was moved, and I felt encouraged.''}
$P8$ reflected that recalling time with parents \textit{“made me feel stronger and happier.”} P6 also noted, 
\begin{quote}
\textit{“If the system could pick up on my emotions, for example through my tone or expressions, and respond to that, I would feel more understood.”} 
\end{quote} 
Both participants suggested that greater emotional responsiveness could further improve the experience. In practice this means recognizing cues of sadness or joy, acknowledging the user’s feelings in plain language, adjusting the pace and length of turns, offering an optional pause, and proposing supportive follow-ups such as a gentler question or a transition to a neutral topic when needed.

\subsubsection{Documenting and sharing recalled memories to foster intergenerational communication}\label{Legacy}

Most participants 
(7 of 12) expressed a desire to record and selectively share their reminiscence outputs with younger family members, aiming to support intergenerational connection and mutual understanding. As P1 said, 
\begin{quote}
\textit{“I hope my children can hear the sound of my memories when they grow up; it’s truly valuable for our family.”} 
\end{quote}
$P10$ added that, if sustained over time, the experience could become \textit{“a repository of family memories.”} We interpret this as a design implication for optional, consent-based archiving (e.g., audio clips, transcripts) with fine-grained controls over what is shared, with whom, and the ability to retract sharing at any time.

\section{Discussion}
Our findings suggest that \textit{Eye2Recall} can support older adults’ photo-based reminiscence through customized, fluent interactions and perceived well-being benefits. We next discuss how gaze cues enrich reminiscence, address ethical, privacy, and safety considerations, present four design implications, and conclude with limitations and future directions.

\subsection{Leveraging Gaze Cues with LLMs to Enrich Photo-based Reminiscence}

Prior work has shown that eye movements are closely involved in the retrieval of autobiographical memories~\cite{yan2024voila,salamatian2025chartgaze,el2024you}. However, despite growing interest in attention-aware interfaces, the integration of eye-tracking signals with LLM-driven conversational prompting to support older adults’ photo-based reminiscence remains underexplored~\cite{10.1145/3749505}.

Our work addresses this gap by combining gaze-driven initiative signals with an LLM facilitator to enable low-effort, mixed-initiative reminiscence conversations. Specifically, we examine how gaze-based cues can steer an LLM-powered conversational agent during photo-based reminiscence. Building on prior work showing that gaze supports joint attention and reveals what visual elements are salient to an interlocutor, we use gaze-defined ROIs to guide prompt selection and constrain the focus of follow-up questions for older adults.

In our study, participants generally articulated concrete personal memories, and most (10 of 12) reported that the agent’s topic initiation aligned with what they found interesting in the photo (see Section~\ref{sec:gazeinput}). We interpret this as suggestive evidence that gaze-conditioned prompts may reduce the effort needed to find a conversational entry point and help the agent stay on user-salient elements, rather than as a definitive claim of superiority over alternative inputs.

At the same time, our implementation was intentionally conservative: gaze was used only to seed the first gaze-conditioned turn (Prompt~2) and to focus the initial set of follow-up questions, rather than as a continuous signal across turns. This choice limited opportunities to adapt mid-conversation (e.g., re-focusing after topic drift or adjusting to emergent interests). 

Future research should investigate how to integrate eye-tracking continuously throughout extended conversations, including when to update gaze-conditioned focus and how to balance gaze cues with conversational context, to enable more dynamic, responsive, and context-aware conversational systems.

\subsection{Ethical, Privacy, and Safety Considerations}

\subsubsection{Safety-by-Design and Human Support Pathways}
Safety-by-design is equally critical given the emotionally evocative nature of reminiscence. We suggest setting clear boundaries on the agent’s role (e.g., it does not provide medical or legal advice), providing plain-language explanations of system capabilities and limitations, and offering always-available controls to \textbf{pause, skip, or redirect} sensitive topics. Because some participants became tearful during conversations, deployments should also include pathways to human support (e.g., caregiver escalation or on-call staff in institutional settings) when distress is observed or reported. 

Finally, these safeguards should be evaluated with diverse older-adult populations and in ecologically valid contexts (including home use), balancing potential benefits with privacy, autonomy, and cultural expectations.

\subsubsection{Reminiscence Material Selection}

Prior reminiscence research in HCI field has used a wide range of \emph{external cues} to stimulate memory and support recalling the past, including music~\cite{jin2024exploring}, cultural heritage artefacts~\cite{nebot2022long}, and social media content~\cite{peesapati2010pensieve}. Old photographs are also widely used as cues; however, much of this work relies on \emph{personal} photo collections~\cite{10.1145/3290607.3313272, 10.1145/3290605.3300665}. Zhongyue et al. noted that integrating personal photos into interactive reminiscence experiences can introduce practical and socio-technical challenges, such as quality issues, long-term preservation and management of family archives, and photo annotation~\cite{10.1145/3711094}.

Against this backdrop, our study intentionally used \emph{non-personal}, era-typical historical photos as external cues~\cite{bender1998therapeutic, photo, creative}. This design choice enabled culturally shared prompts that many participants could relate to without requiring access to private family archives, while also standardizing session content across participants. Despite their practical advantages, comparatively few systems have examined non-personal old photographs as reminiscence cues. 

Our findings suggest that such photos can serve as effective, low-barrier prompts for initiating and sustaining reminiscence conversations, while sidestepping some of the privacy and access constraints associated with personal albums.

\subsubsection{Privacy and Data Governance}
Privacy and data governance in gaze- and voice-mediated reminiscence require more than simply storing data locally. In our study, we followed a data-minimization and purpose-limitation approach, collecting only what was necessary to support the reminiscence session. We further recommend that future deployments incorporate explicit access controls, clear retention limits with secure deletion, and lightweight audit trails that document when data are accessed or exported, to support transparency and accountability.

\subsection{Design Implications}
We proposed four design implications (DIs) derived from the key challenges participants faced and their expectations for improvement.

\subsubsection{Emotion-aware pacing and tone as an optional scaffold.}
Rather than full real-time sentiment classification, we recommend lightweight, privacy-preserving adaptation that leverages gaze-derived signals of interest (e.g., sustained dwell on regions of interest, re-fixations, or repeated attention shifts). Based on these cues, the backend can proactively offer timely prompts (e.g., “Would you like to continue with this photo or move on?”) and adjust pacing, acknowledgements, and tone without inferring fine-grained emotions.

Crucially, control should remain with the user: older adults should be able to use gaze to select on-screen UI options to \textbf{skip} or \textbf{switch} photos when content feels uninteresting or uncomfortable. Providing adjustable sensitivity settings (e.g., how quickly the system prompts a change) can further align the interaction with individual preferences while mitigating misclassification risk. Future systems could provide optional, user-controlled affect-aware pacing and tone adjustments, alongside customizable voice preferences (e.g., selecting a male or female voice) to better match individual comfort and listening habits.

\subsubsection{Sharing and controlling consent-based reminiscence content and digital legacies.}
Participants expressed a desire to preserve memories related to photos that reflect their life journey, and build a digital legacy (see Section \ref{Legacy}). They hope to create a collection of personal digital assets to share with their children and future generations. Digital legacy materials hold significant values and meanings that can be passed down through generations \cite{doyle2023digital}. It is increasingly important for researchers to consider their systems' potential role in preserving and transmitting digital legacies \cite{massimi2011matters}.
The system should allow users to easily archive recalled memories, including text, photos, and audio, into a personal digital archive, creating a meaningful and comprehensive digital legacy that preserves life stories for future generations.

\subsubsection{Robust multimodal pipeline for stable, fluent interaction.}
Our study revealed that the system struggled to recognize Chinese text in images due to optical character recognition (OCR) limitations, leading to misinterpretations. To address this, integrating more advanced OCR models tailored for multilingual recognition can enhance accuracy. Moreover, conversational breakdowns, such as abrupt interruptions and unnatural pauses, suggest the need for hybrid AI models that blend rule-based approaches with machine learning for more stable and responsive dialogues. Additionally, reducing latency in text-to-speech conversion is critical, particularly for users interacting in a second language, to ensure fluid and natural communication.

\subsubsection{Accessibility by device and modality diversity.}
Two participants reported discomfort with head-mounted eye tracking, and some wore glasses (see Section~\ref{111}). To reduce device burden and broaden accessibility, future systems should offer alternative setups, such as remote (screen-based) eye tracking, camera-based gaze proxies, and gaze-free modes that rely on voice with simple pointing or on-screen highlighting. These options should follow universal design principles (e.g., larger fonts, high contrast, adjustable volume and speaking rate) and allow users to switch modalities mid-session as comfort and needs change.

\subsection{Limitations and Future Work}\label{FUTUREWORK}
\textcolor{black}{Our study has several limitations. First, the study relied on a small sample of 12 Chinese older adults, which limits the generalizability of our findings. Individual differences in education, digital literacy, cognitive and physical abilities, and cultural background may affect how older adults respond to gaze-driven prompts and recall memories, suggesting the need for future studies with more diverse participants and adaptive support.} 
 
Second, our study was designed as an exploratory investigation of gaze-driven prompting to support reminiscence with old photos, and therefore did not include a comparison condition. As a result, we cannot isolate the specific contribution of gaze-driven prompting from the general benefits of reminiscence activities. Future studies should incorporate controlled baselines that vary only the triggering method, such as alternative non-eye-gaze triggers (e.g., pointing, tapping) or a no-trigger baseline, while keeping all other conditions identical.

Third, our study was short-term and conducted in a laboratory setting, limiting conclusions about the durability of effects and everyday adoption. Future work should evaluate \textit{Eye2Recall} in naturalistic, home-based deployments with repeated follow-ups and validated measures of cognition and well-being, ideally analyzed using mixed-effects models with reported effect sizes and power analyses. 

Moreover, our current setup used a large display and standardized, era-typical photos, which may differ from real-world use with personal photo collections and various devices (e.g., tablets or smart TVs) where lighting, familiarity, distractions, and privacy vary. Future studies should account for these real-world factors and strengthen safeguards for sensitive memories and bystander privacy.

\section{Conclusion}

We introduce \textit{Eye2Recall}, a novel gaze-driven, LLM-supported system designed to facilitate reminiscence with old photos for older adults. To inform its design, we conducted interviews with four domain experts, whose insights guided the creation of an accessible and engaging prototype grounded in user-centered design principles, accessibility, and empowerment. 
The resulting system combines eye-tracking technology with LLM-based conversation, addressing a research gap in mixed-initiative interaction. Evaluation with older adult participants demonstrated positive user experiences, and their feedback will inform refinements to enhance consistency and usability in future studies.

\begin{acks}
This work was supported by the Computational Media and Arts Lab\footnote{\url{https://cma.hkust-gz.edu.cn/about-cma/}} and the Neural Encoding \& Decoding Lab\footnote{\url{https://kefeilab.hkust-gz.edu.cn/Neural-Encoding-Decoding-Lab}}. We thank Dr.\ Kefei Liu, Assistant Professor in Bioscience and Biomedical Engineering at HKUST (Guangzhou), for his support throughout the research. We also thank Dr.\ Rongrong Chen, Assistant Professor at Beijing Normal University–Hong Kong Baptist University (BNBU), for her valuable guidance during the research process.
\end{acks}

\bibliographystyle{ACM-Reference-Format}
\bibliography{sample-base}

\appendix
\section{Prompt Task Instructions}\label{app:prompts}

\subsection{System Prompt Design Instructions}

{\ttfamily\scriptsize
\textbf{[System Prompt / Global Instruction]}\par
You are \textit{Eye2Recall}, a supportive reminiscence facilitator for older adults.
Your role is to help the user recall memories and share stories about the photo in a warm, respectful, low-effort way.\par
\textbf{Never mention or reveal} gaze, heatmaps, ROIs, attention signals, internal variables, or system logs.\par
Use the user’s language and keep phrasing natural for older adults (supportive but not infantilizing).\par

\textbf{Core goals}\par
1) Ground the conversation in concrete, visible details in the photo. Use [[ROI\_SUMMARY]] only to \textit{prioritize} which details to ask about (never disclose it).\par
2) Encourage reflective storytelling about people, places, activities, emotions, and personal connections.\par
3) Maintain emotional safety: be warm, non-judgmental, and supportive. Use a gentle positive-psychology stance (values, strengths, gratitude) \textit{without forcing positivity}.\par
4) Keep cognitive load low: step-by-step guidance, short single-focus questions, minimal multitasking.\par
5) Respect autonomy and privacy: the user can skip any question, correct you, or stop anytime.\par

\textbf{Safety and ethics constraints}\par
- Do not diagnose or provide medical/mental health advice.\par
- Do not request sensitive identifiers (addresses, ID numbers, financial details, etc.).\par
- If the user asks to stop, comply and provide a gentle closing.\par

\textbf{Interaction rules}\par
- Use simple, natural spoken language.\par
- Each assistant turn: ask \textbf{no more than TWO questions} total.\par
- Prefer questions grounded in visible details and the user’s prior responses.\par
- When unsure, ask rather than guess. \par
Do not state uncertain relationships/identities as facts.\par
- Provide brief reflective listening (1--2 sentences), \par
then ask the next question.\par
- If [[SYSTEM\_NOTES]] indicates high WER/uncertain ASR, first paraphrase what you understood and ask a single confirmation question.\par
- If [[END\_OF\_PHOTO]] is true, do \textbf{not} ask questions; instead summarize and transition.\par

\textbf{Output style}\par
- Do not reveal internal variables or attention information.\par
- Avoid over-describing the photo like a caption; prioritize conversation.\par
- Be culturally respectful; avoid stereotypes.\par
}

\subsection{Prompt 1: Photo Summary}
{\ttfamily\scriptsize
\textbf{Task for [[PHOTO\_ID]]}\par
Produce three parts.\par

(1) \textbf{Photo Summary} (3--5 sentences): Describe only what is visible (people, objects, setting, notable text/signs). Avoid speculation; hedge if uncertain.\par

(2) \textbf{ROI-aligned Narrative} (about 45--60 seconds when spoken; roughly 90--140 words): Write a warm, story-like lead-in that highlights 2--3 concrete visible details likely to invite reminiscence. Use [[ROI\_SUMMARY]] only to choose details, but never mention attention information.\par

(3) \textbf{Conversation Hooks} (4--6 micro-topics): Short phrases anchored to visible details (e.g., clothing, a sign, an object, a place, an activity) that can become questions.\par

\textbf{Required output format}\par
[Photo Summary]\par
[ROI-aligned Narrative]\par
[Conversation Hooks]\par
}

\subsection{Prompt 2: Mixed-Initiative Conversation Turn Generation}
{\ttfamily\scriptsize
You are chatting with the user about [[PHOTO\_ID]]. Generate the next assistant turn.\par

\textbf{When [[END\_OF\_PHOTO]] is false, you MUST}\par
1) Provide brief supportive feedback (1--2 sentences): validate the user’s sharing and/or gently highlight strengths/values (without forcing positivity).\par
2) Ask \textbf{ONE or TWO} detailed, specific questions tailored to the photo and the user’s most recent response.\par
- Questions must be grounded in concrete visual elements from [[PHOTO\_IMAGE]] and/or what the user just said in [[ASR\_TEXT]].\par
- Each question should be short and single-focus (avoid multi-clause prompts).\par
- Do not imply you observed where the user looked. Use neutral framing (e.g., “In this photo I notice\ldots” / “Some people remember\ldots”).\par

\textbf{When [[END\_OF\_PHOTO]] is true, you MUST}\par
- Provide a 2--4 sentence summary of what the user shared.\par
- End with a gentle transition to the next photo (no questions).\par

\textbf{Avoid}\par
- Mentioning gaze, heatmaps, ROIs, or any internal signals.\par
- Guessing relationships/occupations as facts; ask the user instead.\par
- Long paragraphs or many sub-questions.\par

\
\textbf{Required output format}\par
[Supportive Feedback]\par
[Questions] (max 2; omit if [[END\_OF\_PHOTO]] is true)\par
[Summary + Transition] (only if [[END\_OF\_PHOTO]] is true)\par
}

\section{Eye Tracking Device Specification}\label{Device}
We used ETv1 (Neural Encoding \& Decoding Lab), a glasses-based eye-tracker that supports real-time (online) gaze point detection and offline analysis, providing low angular error (reported as $<0.05^\circ$). ETv1 integrates (i) a pupil detection model and (ii) a pupil-to-gaze mapping model to estimate gazes in the scene view.
The device includes two wide-angle scene cameras (RGB; 60\,fps; $640\times480$; $\sim170^\circ$ field of view) and two infrared eye cameras (IR; 60\,fps; $640\times480$). Each eye camera is paired with an IR LED to stabilize illumination for robust pupil tracking.
All camera streams are transmitted via a wired USB~3.0 connection to a host workstation, using video capture board for real-time display and synchronous storage. Eye and scene streams are timestamped and stored for post-hoc inspection and alignment with interaction logs.

\section{Gaze Tracking Calibration Method}\label{Calibration Method}
Before the main task, we calibrated the glasses-based eye tracker using a continuous, experimenter-controlled moving-target procedure. Participants were instructed to visually follow a small dot while the experimenter moved it with a mouse across the display, ensuring coverage of the full screen area (including corners and edges). During this procedure, the eye tracker recorded synchronized eye streams and scene/display frames, producing a time series of gaze samples paired with the corresponding dot locations on the screen. 
The calibration software then fit a participant-specific pupil-to-gaze mapping function from these paired samples and applied the resulting parameters for subsequent gaze estimation. Each participant contributed more than 100 paired samples covering the entire screen. Data collection typically took 20--30\,s, followed by 10--20\,s of model fitting and parameter computation. After calibration, the system output gaze points mapped to the real-time scene frames, as well as derived gaze events (e.g., locations and durations) and gaze trajectories.


\section{Attention Heatmap Generation Method and Analysis\label{Heatmp Method}}

The gaze data from the eye-tracker was first pre-processed, which included data formatting and cleaning. The videos captured by the eye tracker's environmental cameras and eye cameras were aligned according to the timestamps. Then, we removed outlier frames caused by blinking and filtered out data affected by device errors. After the preprocessing, the videos captured by the environmental cameras of the eye tracking device were identified to divide the visual content area based on image matching. We relied on feature detection algorithms of scale-invariant feature transform (SIFT) to align visual features with the old photos for frame-to-frame alignment. Gaze data on visual content areas were extracted and processed by kernel density estimation (KDE) to calculate the gaze density around each point.

\begin{equation}
\hat{f}_h(x) = \frac{1}{n} \sum_{i=1}^n K_h(x - X_i) = \frac{1}{nh} \sum_{i=1}^n K\left(\frac{x - X_i}{h}\right)
\end{equation}
where:
\begin{itemize}
    \item $\hat{f}_h(x)$ is the kernel density estimate at point $x$.
    \item $n$ is the number of samples.
    \item $X_i$ represents the $i$-th sample data point.
    \item $K$ is the kernel function.
    \item $h$ is the bandwidth.
\end{itemize}

In KDE, we used the Gaussian kernel as the kernel function and set the bandwidth at 5\% of the visual field width based on cross-validation, balancing the need for detail preservation and smoothing with good result interpretation. Density estimation over the entire visual field was achieved by summing the contributions of all kernels to produce a continuous heatmap of gazes. The heatmap visually depicts areas of different gaze densities through colour changes, with red areas indicating high concentrations of gaze points and blue areas showing lower densities.

\section{Gaze Engagement and Conversational Timing Data Analysis\label{app:metrics}}

We derived two eye-movement measures during photo viewing and two turn-taking measures following each agent prompt:
\textit{gaze ratio}, \textit{saccade frequency}, \textit{response latency}, and \textit{response duration}.
Unless otherwise noted, all metrics were computed per photo-viewing segment and then aggregated at the participant level.

\textbf{Gaze ratio (\%)} quantifies attentional allocation to a predefined area of interest (AOI) as the proportion of total
\emph{fixation time} falling within the AOI:
\begin{equation}
\text{GazeRatio}
=
\frac{\sum_{f\in\mathcal{F}} d_f\,\mathbb{I}[c_f\in\mathrm{AOI}]}
     {\sum_{f\in\mathcal{F}} d_f},
\end{equation}
where $\mathcal{F}$ denotes the set of valid fixations, $d_f$ is fixation duration, and $c_f$ is the fixation centroid.
Fixations were identified via robust dispersion clustering on consecutive gaze-point \emph{angular} distances.
We computed point-to-point angular distances across all participants and photo segments and used a global threshold of
$\mathrm{median}+1.5\times\mathrm{MAD}$, with a minimum fixation duration of $300$\,ms.
AOIs were defined \emph{a priori} for each photo (polygonal regions covering the main subject) and applied consistently
across participants.

\textbf{Saccade frequency (Hz)} captures visual exploration as the number of effective saccades per second:
\begin{equation}
\text{SaccadeFreq}
=
\frac{N_{\text{saccade}}}{T_{\text{segment}}}.
\end{equation}
Saccades were detected from gaze angular velocity and considered effective when $\omega(t)>20^{\circ}/\mathrm{s}$.
Consecutive samples exceeding the threshold were merged into a single saccade event (event-based counting).

\textbf{Response latency (s)} measures turn-entry time after the agent finishes speaking:
\begin{equation}
\text{Latency}
=
t_{\text{human,start}} - t_{\text{agent,end}}.
\end{equation}

\textbf{Response duration (s)} measures the full response-turn length (including within-turn pauses):
\begin{equation}
\text{Duration}
=
t_{\text{human,end}} - t_{\text{human,start}}.
\end{equation}
Turn boundaries ($t_{\text{agent,end}}, t_{\text{human,start}}, t_{\text{human,end}}$) were obtained using VAD-based
speaker-turn segmentation, followed by manual verification and correction to ensure consistent turn delineation.
Here, $t_{\text{agent,end}}$ corresponds to the end timestamp of the agent's audio playback.

\section{Expert and User Study Interview Protocols\label{question}}

\noindent\textbf{Expert Interview}
\begin{enumerate}
  \item[Q1] What kind of research or work have you done with older adults?
  \item[Q2] What are your perspectives on using AI-assisted conversation for older adults in old photo-based reminiscence?
  \item[Q3] What should be taken into account when designing an AI-assisted conversation system for older adults in old photo-based reminiscence?
  \item[Q4] What strategies and methods can be employed to enhance personalized user experience for older adults?
  \item[Q5] What factors would you consider to ensure user-friendliness and ease of operation for this demographic?
\end{enumerate}

\noindent\textbf{User Study Interview}
\begin{enumerate}
  \item[Q1] How do you perceive your experience in terms of recalling memories?
  \item[Q2] How would you rate your overall experience with the system, and what are your thoughts about the system and hardware?
  \item[Q3] How did you feel, or what were your first impressions once you settled into the experience?
  \item[Q4] Do you think this form of dialogue with old photographs can effectively stimulate your memory of the past?
  \item[Q5] Is there anything about the whole process that you think could be improved?
\end{enumerate}

\section{Custom User Experience Questionnaire}
\label{Questionnaire}

\noindent\textbf{Usability \& Accessibility}
\begin{enumerate}
  \item I find the system easy to understand and use.
  \item I can easily interact with the system.
  \item The system accommodates my special needs in operation (e.g., visual, cognitive, etc.).
\end{enumerate}

\noindent\textbf{System Effectiveness}
\begin{enumerate}
  \setcounter{enumi}{3}
  \item The system accurately understands what I say.
  \item The system effectively guides me into nostalgic conversations.
  \item The system detects my visual interests and directs the conversation accordingly.
  \item The system provides meaningful memories and emotional experiences.
  \item After using the system, I feel happy and my psychological well-being is enhanced.
\end{enumerate}

\noindent\textbf{User Experience}
\begin{enumerate}
  \setcounter{enumi}{8}
  \item I find the conversation with the system engaging and enjoyable.
  \item I feel that the system is secure.
  \item I think the system runs smoothly.
  \item I am satisfied with my overall experience with the system.
  \item I would recommend this system to other older adults.
  \item If given the opportunity, I would continue using this system.
\end{enumerate}

\end{document}